\documentclass[onecolumn,preprint,superscriptaddress]{revtex4-1}
\usepackage{bibentry,natbib}
\usepackage{latexsym,bm}
\usepackage{graphicx}
\usepackage{float}
\usepackage{rotating}
\usepackage{amsmath}
\usepackage{rotating}
\usepackage{color}
\usepackage[colorlinks=true,linkcolor=blue,citecolor=blue,urlcolor=blue]{hyperref}
\usepackage{subfigure}
\usepackage{ulem}
\DeclareGraphicsExtensions{.jpg}

\begin{document}
	\title{Positronium formation and threshold behavior in positron-sodium collisions at low energies}
	\author{Ning-Ning Gao}
	\affiliation{Wuhan Institute of Physics and Mathematics, Innovation Academy for Precision Measurement Science and Technology, Chinese Academy of Sciences, Wuhan 430071, People's Republic of China}
	\affiliation{University of Chinese Academy of Sciences, Beijing 100049, People's Republic of China}
	\author{Hui-Li Han}
	\email{huilihan@wipm.ac.cn}
	\affiliation{Wuhan Institute of Physics and Mathematics, Innovation Academy for Precision Measurement Science and Technology, Chinese Academy of Sciences, Wuhan 430071, People's Republic of China}
	\author{Ting-Yun Shi}
	\affiliation{Wuhan Institute of Physics and Mathematics, Innovation Academy for Precision Measurement Science and Technology, Chinese Academy of Sciences, Wuhan 430071, People's Republic of China}
	\date{\today}
	
	\begin{abstract}
We investigate the elastic and inelastic scattering of positrons by sodium atoms in both the ground state, Na($3s$), and excited states, Na*($3p$, $4s$, $3d$), using the hyperspherical coordinate method with a model potential to represent the atomic core. The threshold behavior of positronium (Ps) formation cross sections is analyzed as the positron impact energy $E$ approaches zero. Within this framework, we derive a generalized expression for partial-wave Ps-formation cross sections at low positron energies, which applies to both ground-state and excited-state sodium targets. Our results confirm that the total threshold behavior follows the expected power-law dependence: \( \sigma_{\text{Ps}} \propto E^{-1/2} \) for exothermic reactions and \( \sigma_{\text{Ps}} \propto E^{a} \) for endothermic reactions, where \( a > 0 \). Furthermore, we find that Ps-formation cross sections for positron scattering from excited Na states are significantly larger than those from the ground state in the low-energy region. A notable enhancement of Gailitis-Damburg oscillations is observed above the Ps($n=2$) threshold, which may account for the increase observed in experimental data. Incorporating contributions from excited sodium targets improves agreement with experimental results and may help resolve discrepancies between theoretical predictions and measurements.
	\end{abstract}
	\maketitle
	\section{Introduction}
   In recent decades, positron scattering from alkali atoms has attracted considerable attention from both experimentalists and theoreticians\;\cite{humberston1994,hewitt1993p,ryzhikh1997p,campbell1998p,ke2004optical,anh2005p,surdutovich2002,zhou1994,ryzhikh1997,Ward2012,lugovskoy2012tc,Kadyrov2016review}, primarily due to its intriguing characteristics, such as its low ionization potential and high polarizability. The positronium (Ps) formation channel remains open even at zero incident positron energy, playing a significant role in the collision process. Additionally, alkali atoms possess a relatively simple structure, often approximated as one-electron atoms\;\cite{Mitroy2001,hanbindingenergy2008,Shertzer2006Li,Shertzer2010Na}, further motivating their study. The growing focus on positron scattering is driven by its relevance in various fields and the application of cross-section data in these contexts\;\cite{Charlton1994,Surko2005,cassidyproduction2007,Aguilar2013,Charlton2015}.

Positron scattering from atomic sodium has been extensively studied, particularly in the context of Ps formation. The first measurements of Ps formation cross sections in positron-sodium collisions were reported by Zhou \textit{et al.}\;\cite{zhou1994}. However, only limited agreement was found between these experimental results and earlier theoretical predictions at low energies\;\cite{SGuha1980}. This discrepancy motivated a second set of measurements by Surdutovich \textit{et al.}\;\cite{surdutovich2002}. The two experimental datasets are largely consistent with each other within the bounds of experimental uncertainty and show good agreement with state-of-the-art theoretical calculations for energies above approximately 1 eV\;\cite{hewitt1993p,ryzhikh1997p,campbell1998p,surdutovich2002,ke2004optical,anh2005p}. In contrast, at energies below about 1 eV, the experimental results exhibit a rising trend in the Ps formation cross section\;\cite{zhou1994,surdutovich2002}, while theoretical models predict a decreasing behavior\;\cite{ryzhikh1997p,campbell1998p,anh2005p}. Surdutovich \textit{et al.}\;\cite{surdutovich2002} provide a detailed discussion of potential sources of experimental error and conclude that none of them are sufficient to explain the observed discrepancy with theory.

Notably, two independent close-coupling (CC) calculations by Ryzhikh and Mitroy\;\cite{ryzhikh1997p} and Campbell \textit{et al.}\;\cite{campbell1998p}, as well as a two-center convergent close-coupling (CCC) calculation\;\cite{lugovskoy2012tc}, suggest that Ps-formation cross sections should decrease as the positron energy drops $2$ eV, which contrasts with earlier experimental measurements\;\cite{zhou1994,surdutovich2002}. This low-energy behavior was also been confirmed by the hyperspherical close-coupling method\;\cite{anh2005p}.
Later, Lugovskoy \textit{et al.}\;\cite{lugovskoy2013threshold} examined the threshold behavior of the elastic and Ps-formation cross sections for a zero partial wave over the energy range from 10$^{-5}$ to 1 eV for Li, Na, and K. Their work confirmed the threshold laws proposed by Wigner\;\cite{wigner1948behavior} but could not resolve the discrepancy observed in positron-sodium experiments at low energies.

   A detailed theoretical investigation of threshold features in positron-atom scattering is critical due to the experimental challenges in determining cross sections at threshold energies\;\cite{Ambalampitiya2023,bray2023}. For positron-alkali-metal scattering, the Ps($n=1$) formation threshold lies at zero energy. For such exothermic reactions, the Wigner threshold law predicts that the \(s\)-wave cross section behaves as \(E^{-1/2}\)\;\cite{wigner1948behavior}. This behavior was confirmed in calculations of the \(s\)-wave Ps formation cross section for positron-sodium scattering\;\cite{lugovskoy2013threshold}. In contrast, Ps formation into the $n=2$ state is an endothermic reaction for ground and low-lying excited states of sodium. Consequently, the threshold behavior is expected to differ from that of Ps($n=1$). Additionally, when a long-range effective dipole interaction is present, the threshold behavior is further modified, as shown by Gailitis and Damburg\;\cite{Gailitis1963}. Studies on antihydrogen formation in low-energy antiproton collisions with excited Ps atoms have demonstrated an \(E^{-1}\) dependence of the cross section\;\cite{nature2017,hyperfine2018}.
   Given the significant discrepancies between experimental and theoretical results for Ps-formation cross sections in positron-sodium scattering at low energies, it is crucial to examine the threshold behavior of higher partial waves and to explore the characteristics of Ps-formation cross sections when the sodium atom is in an excited state.

   The goal of this study is to investigate the elastic and inelastic scattering of positrons by sodium atoms in both the ground state Na($3s$) and the excited states, Na*($3p, 4s, 3d$), with a particular focus on predicting the behavior of Ps-formation cross sections as the incident positron momentum approaches zero. We employ the $R$-matrix propagation method in a hyperspherical coordinate framework, using a model potential for the interactions between charged particles. The hyperradius is divided into two regions: At short distances, the slow-variable-discretization (SVD) method\;\cite{Tolstikhin1996SVD} is applied to overcome numerical difficulties at sharp nonadiabatic avoided crossings, while at large distances, the traditional adiabatic hyperspherical method is utilized to avoid the high memory and computational costs associated with SVD. The $\underline{\mathcal{R}}$ matrix is then propagated from short to large distances, and scattering properties are obtained through the $\underline{\mathcal{S}}$ matrix by matching the $\underline{\mathcal{R}}$ matrix with asymptotic functions and boundary conditions\;\cite{WangJia2011}.

   This paper is organized as follows:
   In Sec. II, we provide an overview of the theoretical method and describe the model potentials employed.
   In Sec. III, we discuss the Ps-formation cross sections and their threshold behavior as the incident positron energy approaches zero.
   Finally, we provide a brief summary.
   Atomic units are used throughout this paper unless explicitly stated otherwise.

	\section{Theoretical formalism}
\begin{figure}[htbp]
		\centering
		\subfigure{
			\includegraphics[width=8.0cm]{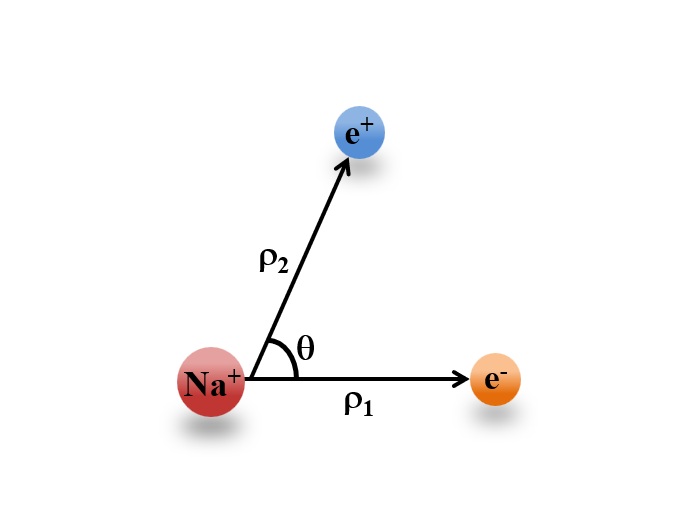}
		}
		\caption{The Jacobi coordinates for the $e^{\scriptscriptstyle+}$-Na system.}
		\label{fig9}
	\end{figure}

	In this study, we investigate the collision properties of the $e^{\scriptscriptstyle+}$-Na system. The sodium atom may be accurately represented as the single valence electron interacting with the core via a local central potential, and the positron-sodium atom system therefore reduces to the equivalent three-body system. The masses of the three charged particles, the Na$^{\scriptscriptstyle+}$ core, $e^{\scriptscriptstyle-}$, and $e^{\scriptscriptstyle+}$, are denoted by $m_{1}$, $m_{2}$, and $m_{3}$, respectively. We employ Delves's hyperspherical coordinates and introduce the mass-scaled Jacobi coordinates. The first Jacobi vector $\vec{\rho}_{\scriptscriptstyle 1}$ is chosen to be the vector from the Na$^{\scriptscriptstyle+}$ core to $e^{\scriptscriptstyle-}$, with reduced mass $\mu_{1}$, and the second Jacobi vector $\vec{\rho}_{2}$ goes from the diatom center of mass to $e^{\scriptscriptstyle+}$, with reduced mass $\mu_{2}$. The angle between $\vec{\rho}_{1}$ and $\vec{\rho}_{2}$ is denoted by $\theta$, as shown in the Figure~\ref{fig9}. The hyperradius $R$ and hyperangle $\phi$ are defined as\\
	\begin{equation}
		\label{1}
		\mu R^{2}=\mu_{1}\rho_{1}^{2}+\mu_{2}\rho_{2}^{2}\,,
	\end{equation}
	and\\
	\begin{equation}
		\label{2}
		\tan\phi=\sqrt{\frac{\mu_{2}}{\mu_{1}}}\frac{\rho_{2}}{\rho_{1}},\;\; 0 \leq\phi\leq\frac{\pi}{2}\,,
	\end{equation}
	respectively, where $R$ is the only coordinate with the dimension of length and represents the overall size of the three-body system. The rotation of the plane containing the three particles is described collectively by $\Omega$ $[\Omega \equiv (\theta, \phi, \alpha, \beta, \gamma)]$, which includes $\theta$, $\phi$, and three Euler angles $(\alpha, \beta, \gamma)$. The parameter $\mu$ is an arbitrary scaling factor, and we choose $\mu=\sqrt{\mu_{1}\mu_{2}}$ for our calculations.
	
The Schr$\mathrm{\ddot{o}}$dinger equation in hyperspherical coordinates can be written after rescaling the three-body wave function $\Psi_{\upsilon'}$ as $\psi_{\upsilon'}(R;\theta,\phi)=\Psi_{\upsilon'}(R;\theta,\phi) R^{5/2} \sin\phi \cos\phi$:\\
	\begin{equation}
		\label{3}
		\left[ -\frac{1}{2\mu}\frac{d^{2}}{dR^{2}}+\left( \frac{\Lambda^{2}-\frac{1}{4}}{2\mu R^{2}}+V(R;\theta,\phi)\right) \right] \psi_{\upsilon'}(R;\Omega)=E\psi_{\upsilon'}(R;\Omega)\,,
	\end{equation}
	where $\Lambda^{2}$ is the square ``grand angular momentum operator", and its expression is as given in Ref.\;\cite{lin1995}. The three-body interaction potential $V(R;\theta,\phi)$ is expressed as:\\
	\begin{equation}
		\label{4}
		V(R;\theta,\phi)=v_{\scriptscriptstyle \scriptscriptstyle-}(r_{12})+v_{\scriptscriptstyle +}(r_{ 13})+v(r_{23})\,,
	\end{equation}
	where $r_{12}$, $r_{13}$, and $r_{23}$ are the electron-core distance, the positron-core distance, and the electron-positron distance, respectively.
	
For the $e^{\scriptscriptstyle+}$Na system, the model potential used for the valence electron and the core is the same as that used by Han \textit{et al.}\;\cite{hanbindingenergy2008} and has the form
	\begin{equation}
		\label{5}
		v_{\scriptscriptstyle-}(r_{12})=-\frac{1}{r_{12}}\left[Z_{c}+(Z-Z_{c})e^{-a_{1}r_{ 12}}+a_{2}r_{12}e^{\scriptscriptstyle -a_{3}r_{12}}\right]-\frac{a_{c}}{2r_{12}^{4}}\left[W_{3}\left(\frac{r_{12}}{r_{c}}\right)\right]^{2}\,,
	\end{equation}
where
	\begin{equation}
	\label{6}
     W_{n}(r)=1-e^{-r^{n}}\,,
     \end{equation}
is the cutoff function used to ensure the correct behavior at the origin. The second term in Eq.\;(\ref{5}) describes the core polarization, where $a_{c}$ = 0.9457 a.u. is the Na$^{\scriptscriptstyle+}$ polarizability\;\cite{Johnson1983}. The nuclear charge is $Z=11$, and the charge of the Na$^{\scriptscriptstyle+}$ core is $Z_{c}=1$. The remaining parameters, which are $a_{1}=3.32442452,\, a_{2}=0.71372798,\, a_{3}=1.83281815$, and $r_{c}=0.52450638$, are fitted from Ref.\;\cite{Liu1999} using the least-squares method to reproduce the experimentally measured energy levels of the Na atom. For the positron-core interaction $v_{\scriptscriptstyle +}(r_{13})$, the first term is the same as in $v_{\scriptscriptstyle-}(r_{12})$, but with the opposite sign, and the polarization potential is chosen to be the same as in $v_{\scriptscriptstyle -}(r_{12})$. The electron-positron interaction $v(r_{23})$ is given by
\begin{equation}
	\label{7}
v(r_{23})=-\frac{1}{|\vec{r}_{13}-\vec{r}_{12}|}+\frac{a_{c}}{r_{12}^{ 2}r_{13}^{2}}\cos{\theta} W_{ 3}\left(\frac{r_{12}}{r_{c}}\right)W_{3}\left(\frac{r_{13}}{r_{c}}\right)\,,
\end{equation}

	The three-body wave function $\psi_{\upsilon'}$ can be expanded with the complete, orthonormal set of the angular wave function $\Phi_{\nu}$ and radial wave functions $F_{\nu\upsilon'}$ as
	\begin{equation}
		\label{8}
		\psi_{\upsilon'}(R;\Omega)=\sum\limits_{\nu=0}^{\infty}F_{\nu\upsilon'}(R)\Phi_{\nu}(R;\Omega)\,.
	\end{equation}
	The adiabatic potentials $U_{\nu}(R)$ and channel functions $\Phi_{\nu}(R;\Omega)$ at fixed $R$ can be obtained by solving the following adiabatic eigenvalue equation:
	\begin{equation}
		\label{9}
		\left( \frac{\Lambda^{2}-\frac{1}{4}}{2\mu R^{2}}+V(R;\theta,\phi)\right) \Phi_{ \nu}(R;\Omega)=U_{ \nu}(R) \Phi_{\nu}(R;\Omega)\,.
	\end{equation}

The channel function is further expanded on Wigner rotation matrices $D_{KM}^{J}$ as
\begin{align}
	\label{10}
	\Phi_{\nu}^{J\Pi M}(R;\Omega)=\sum\limits_{K=0}^{J}u_{\nu K}(R;\theta,\phi)\overline{D}_{ KM}^{ J\Pi}(\alpha,\beta,\gamma)\,.
\end{align}
\begin{align}
	\label{11}
	\overline{D}_{KM}^{J\Pi}=\frac{1}{4\pi}\sqrt{2J+1}[D_{KM}^{J}+(-1)^{K+J}\Pi D_{-KM}^{J}]\,,
\end{align}
where $J$ is the total nuclear orbital angular momentum, $M$ is its projection onto the laboratory-fixed axis, and $\Pi$ is the parity with respect to the inversion of the nuclear coordinates. The quantum number $K$ denotes the projection of $J$ onto the body-frame $z$ axis and takes the values $K=J,J-2,\ldots,-(J-2),-J$ for the ``parity-favored" case, $\Pi=(-1)^{ J}$, and $K=J-1,J-3,\ldots,-(J-3),-(J-1)$ for the ``parity-unfavored" case, $\Pi=(-1)^{J+1}$. $u_{\nu K}(R;\theta,\phi)$ is expanded with $B$-spline basis sets:

\begin{align}
	\label{12}
	u_{\nu K}(R;\theta,\phi)=\sum\limits_{i}^{ N_{\phi}} \sum\limits_{j}^{N_{\theta}}c_{ i,j}B_{i}(\phi)B_{j}(\theta)\,,
\end{align}
where $N_{\theta}$ and $N_{\phi}$ are the sizes of the basis sets in the $\theta$ and $\phi$ directions, respectively. Using the $B$-splines as a basis function has multiple advantages, including high localization, flexibility, completeness, and numerical stability\;\cite{Bachau2001,Kang2006}, which enable us to obtain accurate potential curves and channel functions by employing these advantages. A detailed investigation of the knot distribution suitable for calculations of the positron alkali-atom bound states can be found in our previous work\;\cite{hanbindingenergy2008}.

The goal of our scattering study is to determine the scattering matrix $\underline{\mathcal{S}}$ from the solutions of Eq.\;(\ref{3}). We first calculate the $\underline{\mathcal{R}}$ matrix, which is defined as
\begin{align}
\label{13}
\underline{\mathcal{R}}(R)=\underline{\textsl{F}}(R)[\widetilde{\underline{\textsl{F}}}(R)]^{ -1}\,,
\end{align}
where matrices $\underline{\textsl{F}}$ and $\widetilde{\underline{\textsl{F}}}$ can be calculated from the solution of Eqs.\;(\ref{3}) and (\ref{9}) by evaluating the integrals:
\begin{align}
\label{14}
F_{\nu,\upsilon'}(R)=\int d\Omega \Phi_{\nu}(R;\Omega)^{\scriptscriptstyle *}\psi_{ \upsilon'}(R;\Omega)\,,
\end{align}
\begin{align}
\label{15}
\widetilde{F}_{\nu,\upsilon'}(R)=\int d\Omega \Phi_{\nu}(R;\Omega)^{\scriptscriptstyle *}\frac{\partial}{\partial R}\psi_{\upsilon'}(R;\Omega)\,.
\end{align}
The hyperradius $R$ is divided into $(N-1)$ intervals using a set of grid points $R_{ 1}<R_{ 2}<\cdots R_{ N}$. At short distances, we utilize the SVD method to solve Eq.\;(\ref{3}) in the interval $[R_{ i},R_{i+1}]$. In this method, we express the total wave function $\psi_{\upsilon'}(R;\Omega)$ in terms of the discrete variable representation (DVR) basis $\pi_{i}$ and the channel functions $\Phi_{\nu}(R;\Omega)$ as follows:
\begin{align}
\label{16}
 \psi_{\upsilon'}(R;\Omega)=\sum^{ N_{ DVR}}_{i}\sum^{ N_{chan}}_{\nu}C^{ \upsilon'}_{ i\nu}\pi_{ i}(R)\Phi_{\nu}(R_{i};\Omega)\,,
\end{align}
where $N_{DVR}$ represents the number of DVR basis functions and $N_{chan}$ is the number of included channel functions.
By inserting $\psi_{ \upsilon'}(R;\Omega)$ into the three-body Schr$\mathrm{\ddot{o}}$dinger equation given by Eq.\;(\ref{3}), we arrive at a standard algebraic problem for the coefficients $C^{\upsilon'}_{i\nu}$:
\begin{align}
\label{17}
 \sum^{N_{DVR}}_{j}\sum^{ N_{chan}}_{\mu}\mathcal{T}_{ij}
 \mathcal{O}_{i\nu,j\mu}C^{\upsilon'}_{ j\mu}+U_{\nu} (R_{i})C^{\upsilon'}_{i\nu}=E^{ \upsilon'}C^{ \upsilon'}_{i\nu}\,,
\end{align}
where
\begin{align}
\label{18}
\mathcal{T}_{ij}=\frac{1}{2\mu}\int^{R_{i+1}}_{ R_{ i}}\frac{d}{dR}\pi_{i}(R)\frac{d}{dR}\pi_{j}(R)dR\,,
\end{align}
are the kinetic-energy matrix elements, with ${R_{i}}$ and ${R_{i+1}}$ being the boundaries of the calculation box, and
\begin{align}
\label{19}
\mathcal{O}_{i\nu,j\mu}=\langle\Phi_{\nu}(R_{i};\Omega)|\Phi_{ \mu}(R_{j};\Omega)\rangle
\end{align}
are the overlap matrix elements between the adiabatic channels defined at different quadrature points.

At large distances, the traditional adiabatic hyperspherical method is used to solve Eq.\;(\ref{3}).
When substituting the wave functions $\psi_{\upsilon'}(R;\Omega)$ into Eq.\;(\ref{3}), one obtains a set of coupled ordinary differential equations:
\begin{align}
\label{20}
\left[-\frac{1}{2\mu}\frac{d^{2}}{dR^{2}}+U_{\nu}(R)- E\right]F_{\nu,\upsilon'}(R)
-\frac{1}{2\mu}\sum_{\mu}\left[2P_{\mu\nu}(R)\frac{d}{dR}+Q_{\mu\nu}(R)\right]F_{\mu \upsilon'}(R)=0\,,
\end{align}
where
\begin{align}
\label{21}
P_{\mu\nu}(R)=\int d\Omega \Phi_{ \mu}(R;\Omega)^{\scriptscriptstyle *}\frac{\partial}{\partial R}\Phi_{\nu}(R;\Omega)\,,
\end{align}
and
\begin{align}
\label{22}
Q_{\mu\nu}(R) = \int d\Omega \Phi_{ \mu}(R;\Omega)^{\scriptscriptstyle *}\frac{\partial^{2}}{\partial R^{2}}\Phi_{\nu}(R;\Omega)\,.
\end{align}
are the nonadiabatic couplings that control the inelastic transitions as well as the width of the resonance supported by  adiabatic potential $U_{\nu}(R)$ .

The effective hyperradial potentials that include hyperradial kinetic energy contributions with the $P_{ \nu\nu}^{2}$ term are more physical than adiabatic hyperpotentials and are defined as
\begin{align}
\label{23}
W_{\nu \nu}(R)=U_{ \nu}(R)-\frac{\hbar^{2}}{2 \mu} P_{\nu \nu}^{2}(R)\,.
\end{align}

The effective potentials for the bound-state channels exhibit the following asymptotic behavior at large \( R \):
\begin{equation}
	\label{24}
	W_{\nu \nu}(R) = \frac{l_{\nu} (l_{\nu} +1)}{2\mu R^{2}} + E_{2b}\,,
\end{equation}
where \( E_{2b} \) is the two-body bound-state energy, and \( l_{\nu} \) represents the angular momentum of the third particle relative to the two-body bound system, whose angular momentum is denoted by \( l_{1} \).

In the hyperspherical method, the total angular momentum \( J \), along with \( l_{1} \) and \( l_{\nu} \), follows the selection rule:
\begin{equation}
	\label{25}
|l_{1} - l_{\nu} | \leq J \leq l_{ 1} + l_{\nu}\,.
\end{equation}

Finally, we use the $R$-matrix propagation method. Within an interval $[R_{1},R_{2}]$, given an $\underline{\mathcal{R}}$ matrix at $R_{1}$, we calculate the corresponding $\underline{\mathcal{R}}$ matrix at another point $R = R_{2}$ using
\begin{align}
\label{26}
\underline{\mathcal{R}}(R_{2})=\underline{\mathcal{R}}_{22}
-\underline{\mathcal{R}}_{21}\left[\underline{\mathcal{R}}_{11}
+\underline{\mathcal{R}}(R_{1})\right]^{-1}\underline{\mathcal{R}}_{12}\,.
\end{align}
where the corresponding matrices give:
\begin{align}
	\label{27}
	(\underline{\mathcal{R}}_{11})_{\nu\mu} = \sum\limits_{n }\frac{u_{ \nu}^{(n)}(R_{1})u_{ \mu}^{(n)}(R_{1})}{2\mu (\varepsilon_{n}-E)}\,,
\end{align}
\begin{align}
	\label{28}
	(\underline{\mathcal{R}}_{12})_{\nu\mu} = \sum\limits_{n }\frac{u_{\nu}^{(n)}(R_{ 1})u_{\mu}^{(n)}(R_{2})}{2\mu (\varepsilon_{n}-E)}\,,
\end{align}
\begin{align}
	\label{29}
	(\underline{\mathcal{R}}_{21})_{\nu\mu} = \sum\limits_{n }\frac{u_{\nu}^{(n)}(R_{2})u_{ \mu}^{(n)}(R_{1})}{2\mu (\varepsilon_{n}-E)}\,,
\end{align}
\begin{align}
	\label{30}
	(\underline{\mathcal{R}}_{22})_{\nu\mu} = \sum\limits_{n }\frac{u_{\nu}^{(n)}(R_{2})u_{\mu}^{(n)}(R_{2})}{2\mu (\varepsilon_{n}-E)}\,,
\end{align}
where $\nu$ and $\mu$ denote different channels, indices 1 and 2 do not label the channel, and more details can be found in Ref.\;\cite{WangJia2011}.

The $\underline{\mathcal{K}}$ matrix can be expressed as the following matrix equation:
\begin{align}
\label{31}
\underline{\mathcal{K}}=
(\underline{f}-\underline{f}'\underline{\mathcal{R}})
(\underline{g}-\underline{g}'\mathcal{R})^{-1}\,,
\end{align}
where $f_{\nu \nu'}=\sqrt{\frac{2\mu k_{\nu}}{\pi}} R j_{l_{\nu}}(k_{\nu} R)\delta_{\nu \nu'}$ and $g_{\nu \nu'}=\sqrt{\frac{2\mu k_{\nu}}{\pi}} R n_{l_{\nu}}(k_{ \nu} R)\delta_{\nu \nu'}$ are the diagonal matrices of energy-normalized spherical Bessel and Neumann functions.
For the bound-state channel, $l_{\nu}$ is the angular momentum of the third atom relative to the dimer and $k_{\nu}$ is given by $k_{\nu}=\sqrt{2 \mu\left(E-E_{2 b}\right)}$.
For the continuous channel, $l_{\nu}=\lambda_{\nu}+3 / 2$, and $k_{\nu}=\sqrt{2 \mu E}$. The scattering matrix $\underline{\mathcal{S}}$ is related to $\underline{\mathcal{K}}$ as follows:
\begin{align}
\label{32}
\underline{\mathcal{S}}=(\underline{1}+i\underline{\mathcal{K}})(\underline{1}-i\underline{\mathcal{K}})^{-1}\,.
\end{align}

The cross-sections are expressed in terms of the $\underline{\mathcal{S}}$ matrix as:
\begin{align}
\label{33}
\sigma_{ij}=\sum\limits_{J,\Pi}\frac{(2J+1)\pi}{k_{ad}^{2}}|S_{ f\leftarrow i}^{J,\Pi}-\delta_{ij}|^{2}\,.
\end{align}
where $i$ denotes the incident channel (the initial state) and $j$ denotes the exit channel (the final state). $k_{ad}=\sqrt{2\mu_{ad}(E-E_{2b})}$ is the atom-dimer wave number, and $\mu_{ad}$ is the atom-dimer reduced mass.
\section{Results and Discussion}
	
\subsection{Ps-formation cross sections }

\begin{figure}[htbp]
	\centering
	\subfigure{
		\includegraphics[width=8.6cm,height=6.2cm]{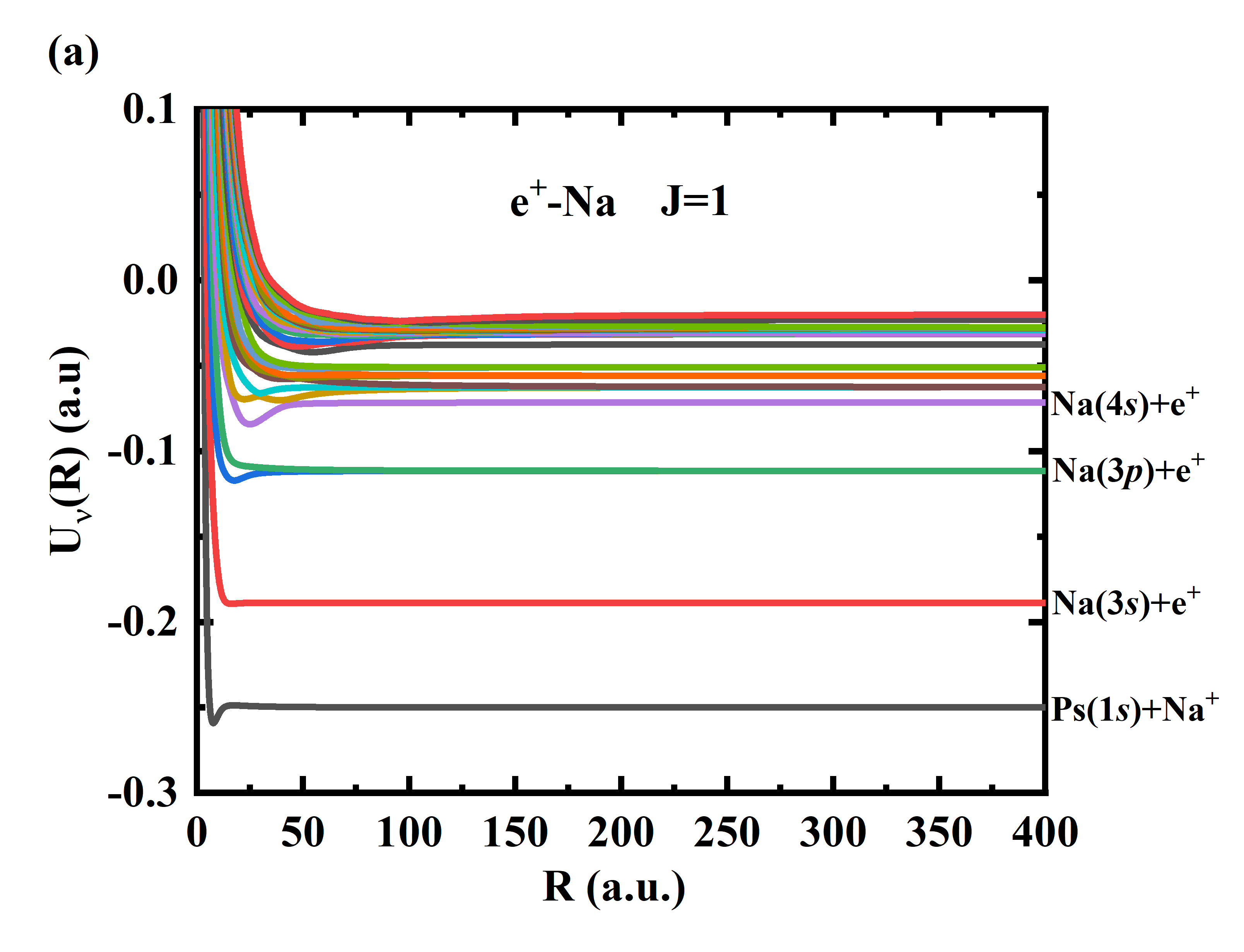}
		\label{fig1a}
	}
	\subfigure{
		\includegraphics[width=7.2cm,height=6.2cm]{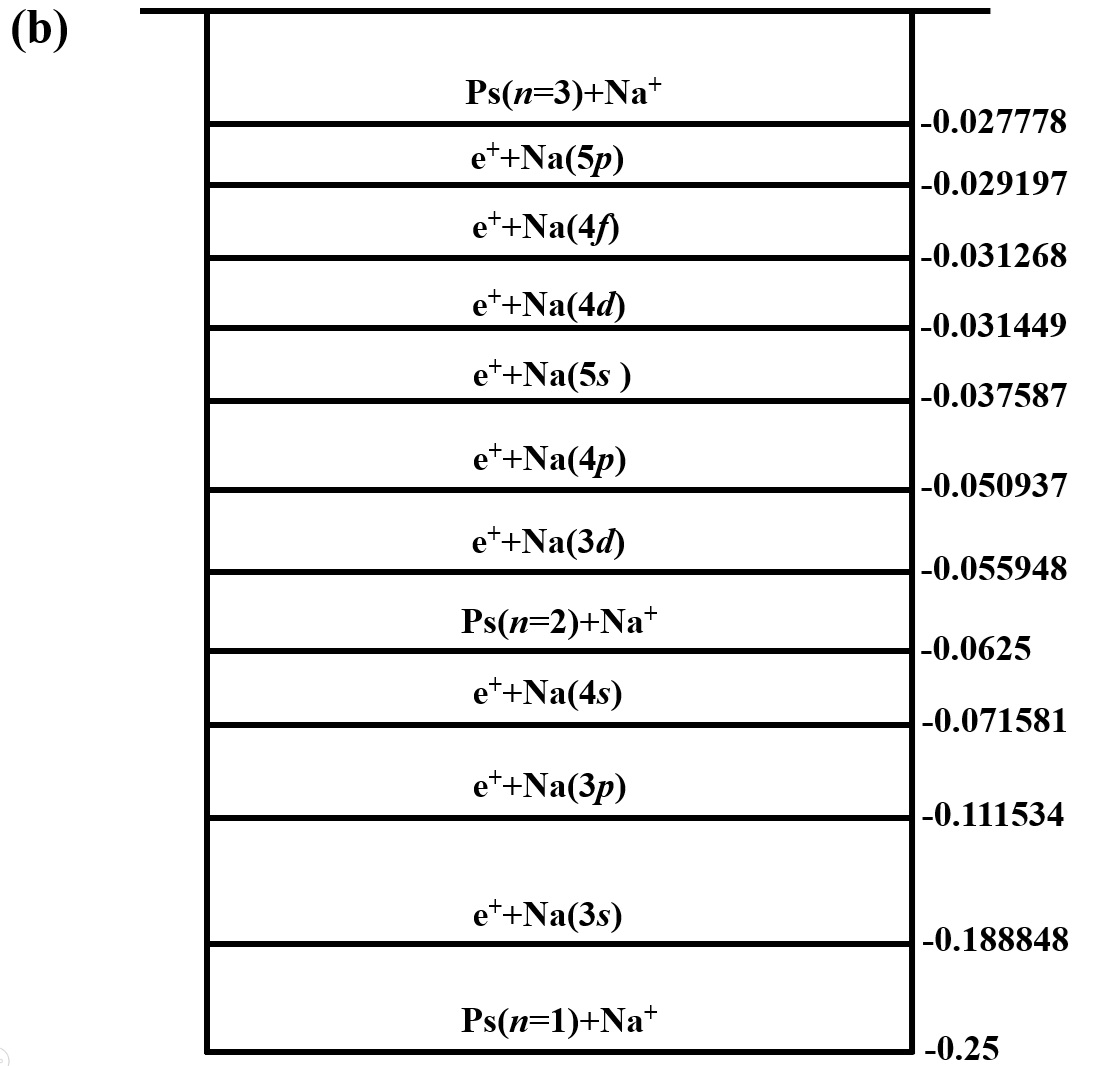}
		\label{fig1b}
	}
\caption{(Color online) (a) The $J=1$ adiabatic hyperspherical potential curves $U_{\nu}$(R) for the $e^{\scriptscriptstyle+}$-Na system. (b) The energy levels (in a.u.) in the $e^{\scriptscriptstyle+}$-Na system.}
\label{fig1}
\end{figure}

Figure~\ref{fig1a} presents the lowest 27 adiabatic potential curves for the $e^{\scriptscriptstyle+}$-Na system with total angular momentum quantum number $J=1$. These calculations employ basis sets with $N_{\theta}=76$ and $N_{\phi}=218$, ensuring accuracy to at least six significant digits. Figure~\ref{fig1a} shows two types of scattering channels: the Ps-formation channel and the atom-plus-positron channel. A key feature of these potentials is the presence of a degenerate state in the Na($3p, 3d$) + $e^{\scriptscriptstyle+}$ channel and an increased number of avoided crossings in higher channels. The lowest channel corresponds to Ps($n=1$) formation, which remains open even at zero scattering energy, making it an exothermic reaction.

Figure~\ref{fig1b} illustrates the energy levels in the $e^{\scriptscriptstyle+}$-Na system, showing that Ps($n=2$) formation is an endothermic reaction when the target sodium atom is in the Na($3s$), Na($4s$), or Na($3p$) state but becomes exothermic when the sodium atom is in the Na($3d$) state.

In this study, we focus on the low-energy behavior of elastic and Ps-formation cross sections over the energy range of $10^{-5}$ to $2.1$ eV, where experimental results exhibit a different energy dependence compared to theoretical predictions. For positron scattering from the Na($3s$) ground state, this energy range includes only two open channels, Ps($1s$) + Na$^{\scriptscriptstyle+}$ and Na($3s$) + $e^{\scriptscriptstyle+}$, for all partial waves, restricting the scattering process to elastic scattering and Ps($n=1$) formation. In contrast, for positron scattering from the excited Na($3p$) state, this energy range corresponds to the gap between Na($5s$) and Na($3p$), where 9 channels are open for $S$ states and 13 are open for nonzero angular momentum states, leading to more complex scattering dynamics. The total Ps-formation cross section accounts for contributions from both the Ps($n=1$) + Na$^{\scriptscriptstyle+}$ and Ps($n=2$) + Na$^{\scriptscriptstyle+}$ channels. For Na($4s$) and Na($3d$) targets, up to 27 channels are open. Due to computational limitations, we calculate Ps-formation cross sections only for energies below 1 eV. Since degenerate channels such as Ps($n \ge 2$) + Na$^{\scriptscriptstyle+}$ and Na($3p, 3d$) + $e^{\scriptscriptstyle+}$ exist, the cross sections are averaged over initial states and summed over final states.

\begin{figure}[htbp]
	\centering
	\subfigure{
		\includegraphics[width=7.8cm]{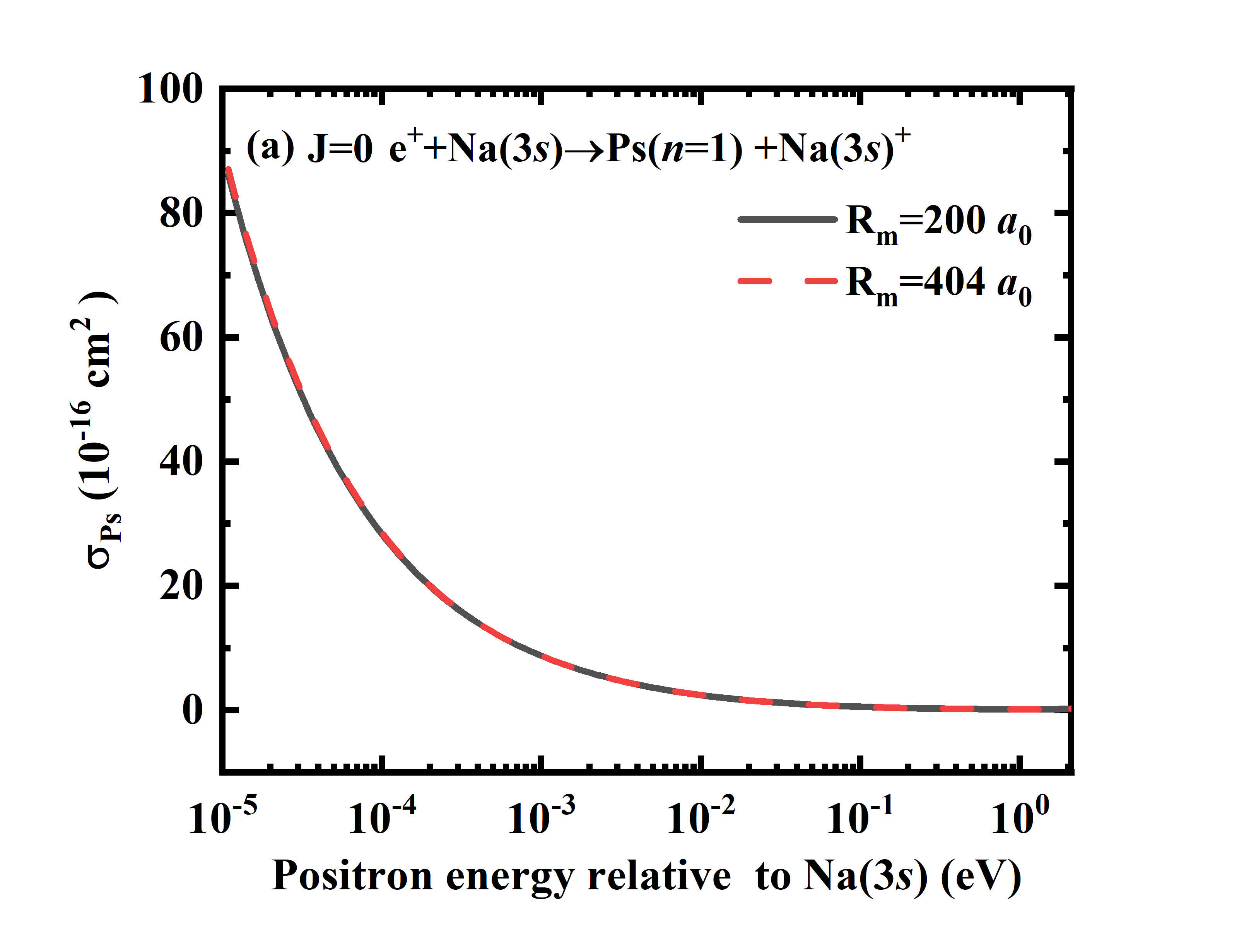}
		\label{fig2a}
	}
	\subfigure{
		\includegraphics[width=7.8cm]{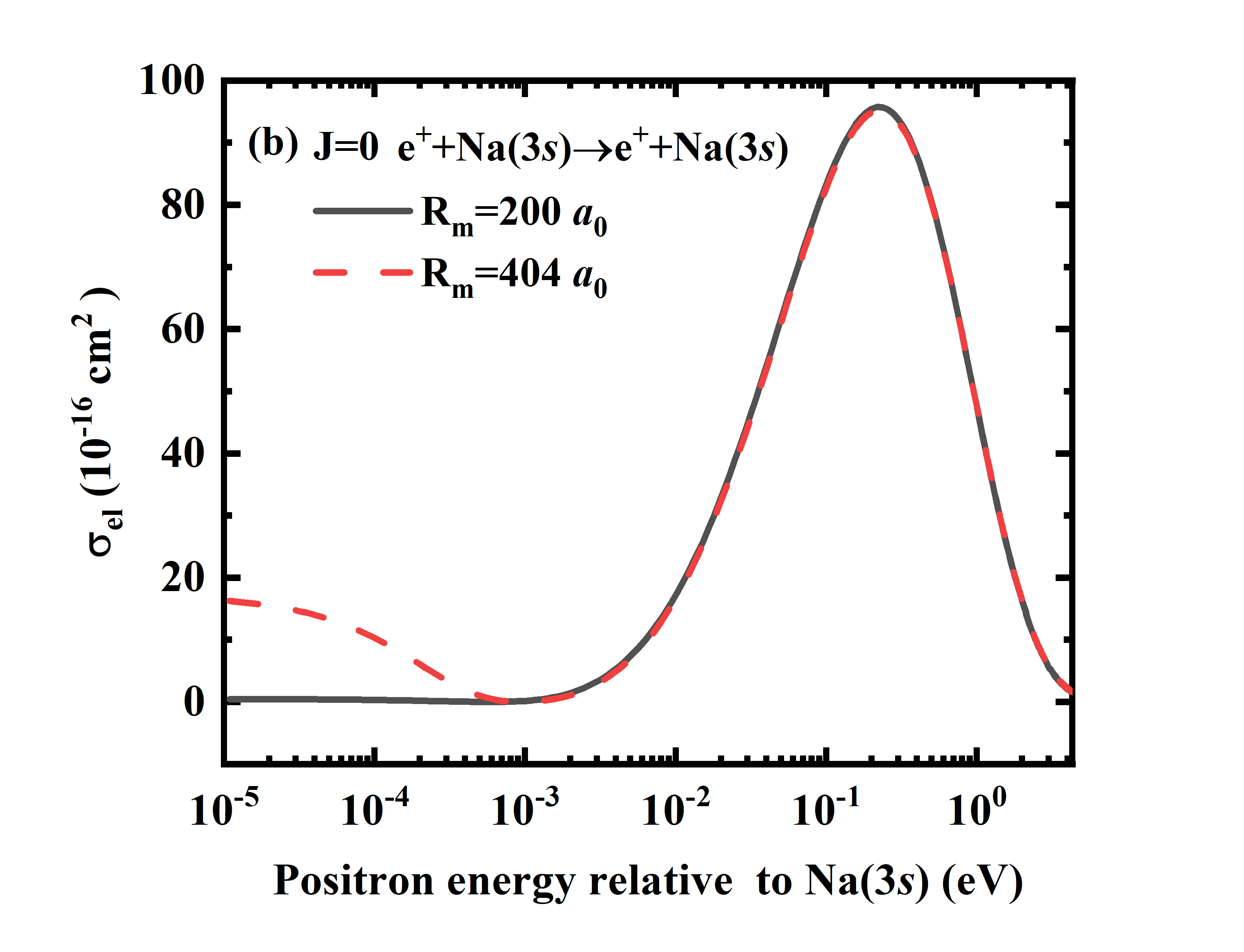}
		\label{fig2b}
	}
	\caption{(Color online) Convergence of (a) the $s$-wave Ps($n=1$)-formation cross sections and (b) the elastic cross sections with respect to the matching distances $R_{m}$ for the collision process of $e^{\scriptscriptstyle+}$ + Na($3s$).}
	\label{fig2}
\end{figure}

 To calculate the Ps-formation cross sections for the $e^{\scriptscriptstyle+}$-Na collision, we employ up to 27 channels. At low energies, only a few partial waves are needed, so we use $J=5$ to obtain the total Ps-formation cross sections. The solutions are matched to asymptotic solutions at different hyperradii, specifically at $R=200 \, a_{\scriptscriptstyle 0}$ and $R=400 \, a_{\scriptscriptstyle 0}$, to ensure the stability of the cross sections with respect to the matching distance. Figures~\ref{fig2a} and~\ref{fig2b} illustrate the convergence of the cross sections with regard to the matching distance for the Ps formation and elastic cross sections with $J=0$, respectively. It is evident that the Ps-formation cross sections are largely insensitive to variations in the matching distance, while the elastic-scattering cross sections exhibit a more pronounced dependence on the matching distance.

\begin{figure}[htbp]
	\centering
	\subfigure{
		\includegraphics[width=7.8cm]{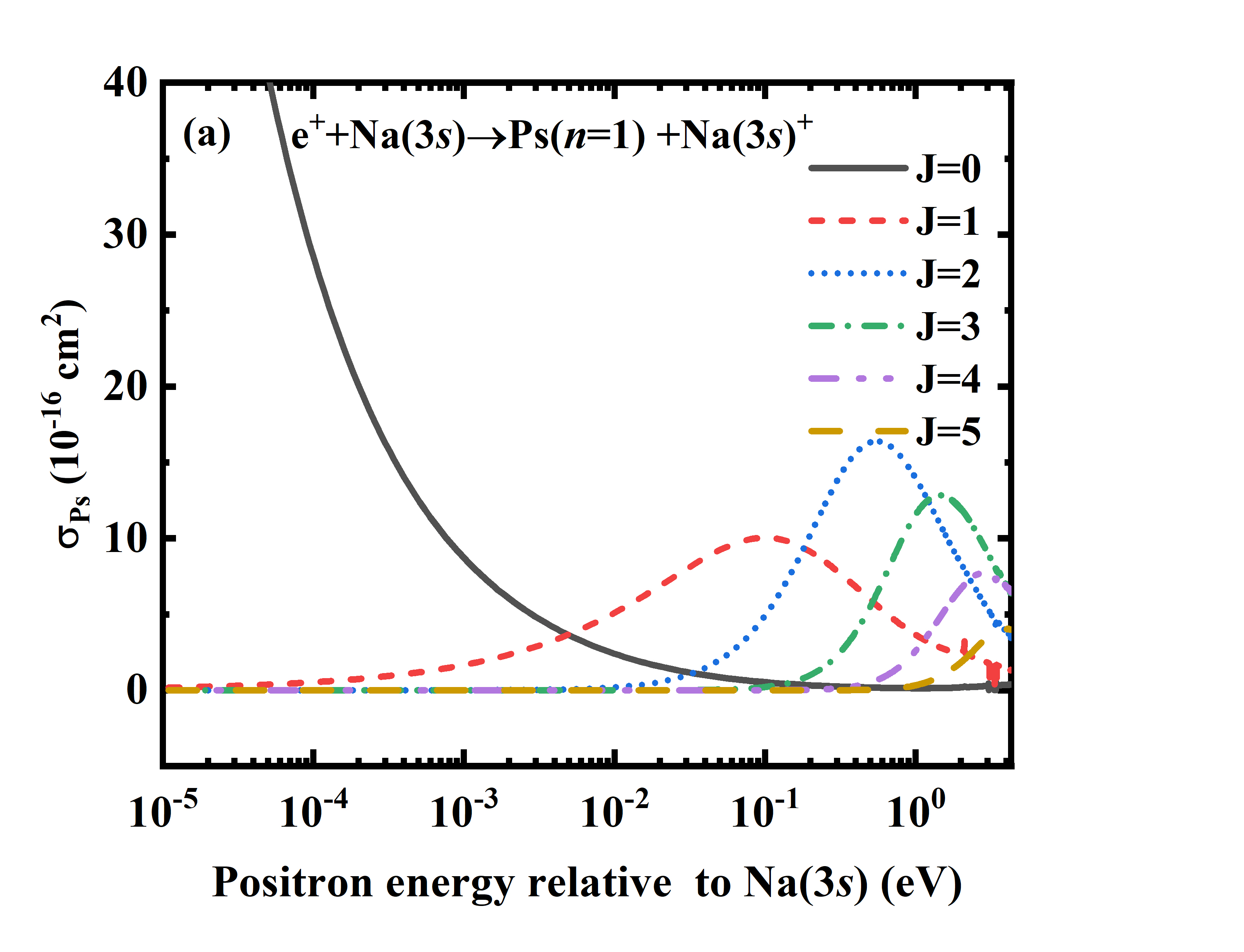}
		\label{fig3a}
	}
	\subfigure{
		\includegraphics[width=7.8cm]{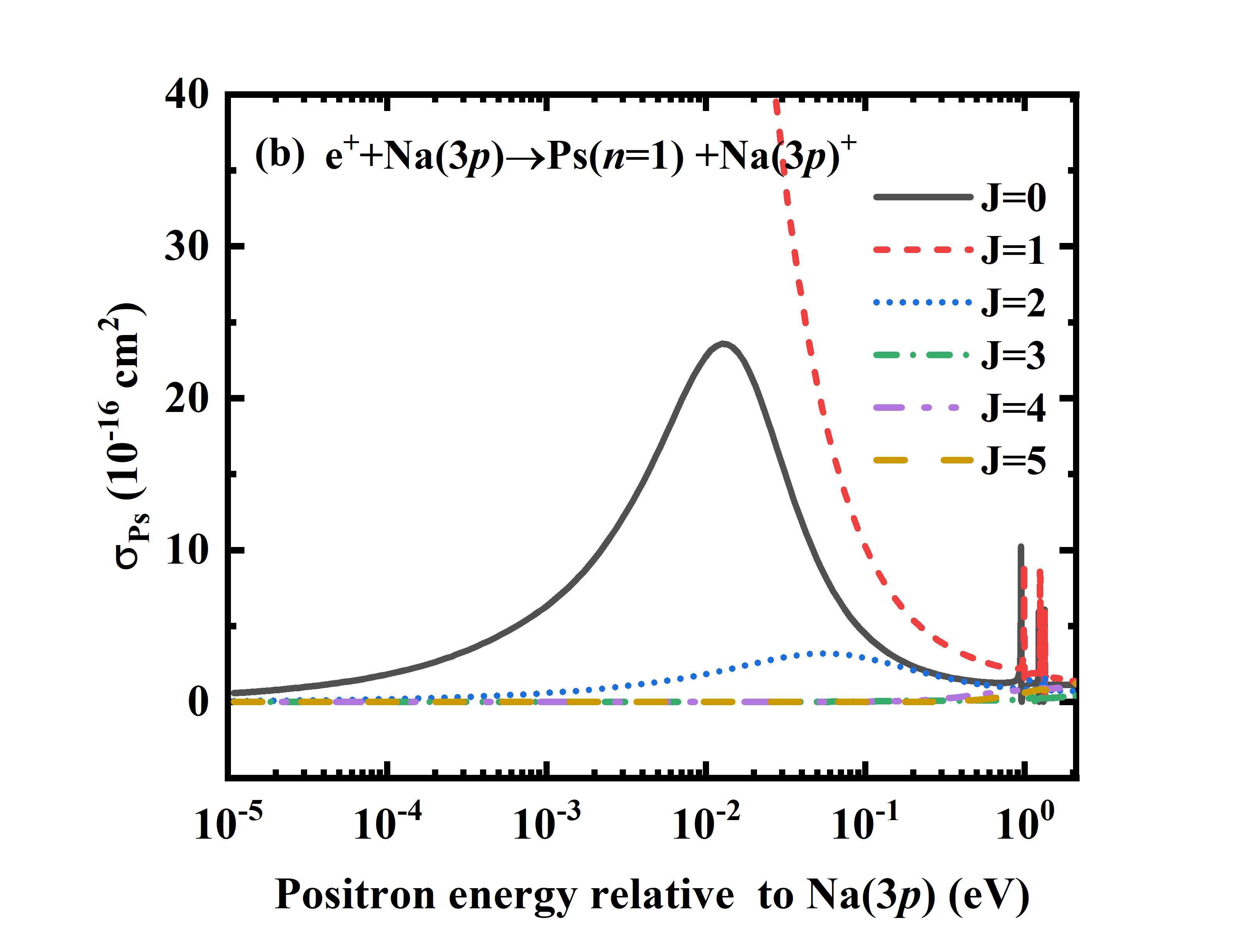}
		\label{fig3b}
	}
    \subfigure{
        \includegraphics[width=7.8cm]{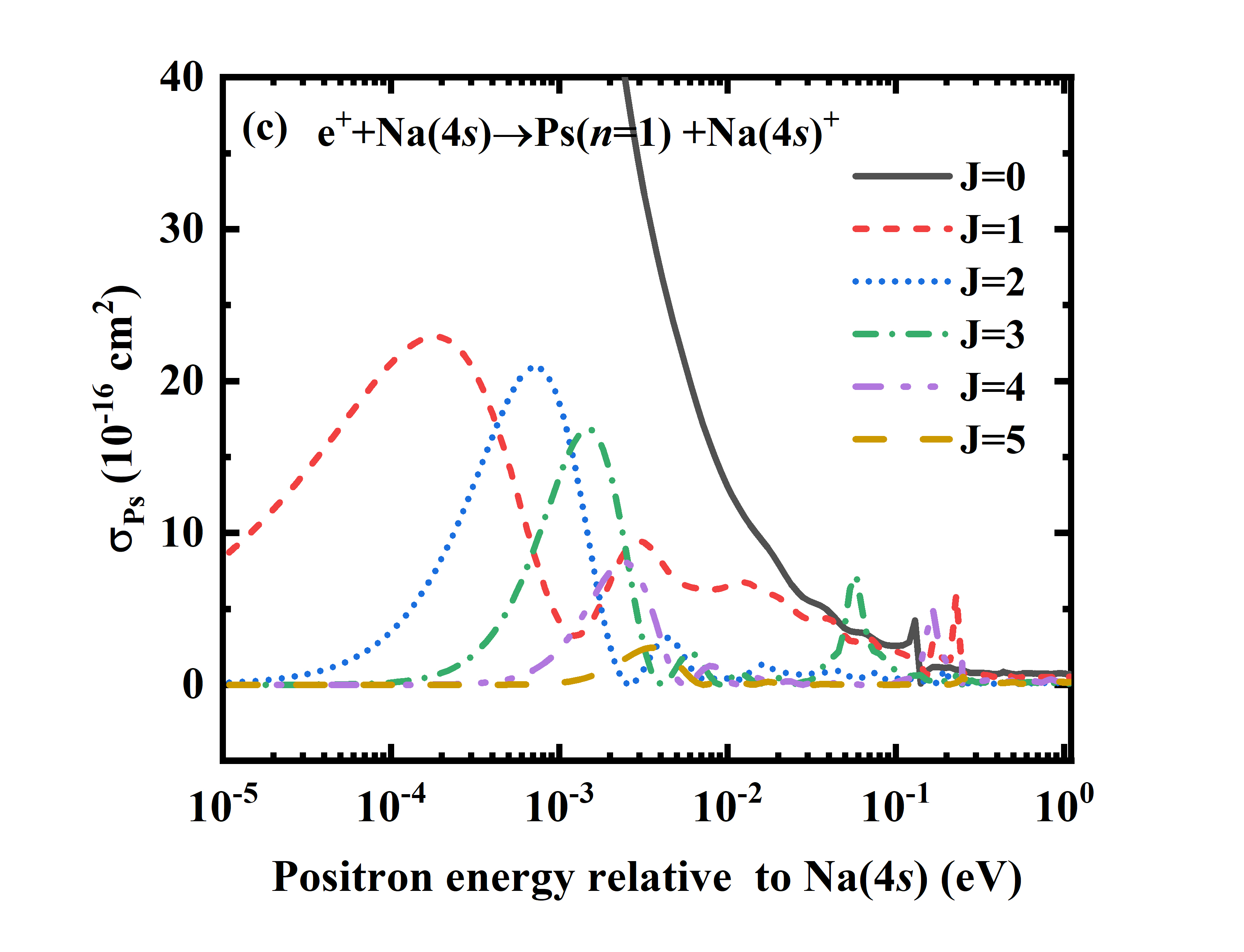}
		\label{fig3c}
	}
    \subfigure{
        \includegraphics[width=7.8cm]{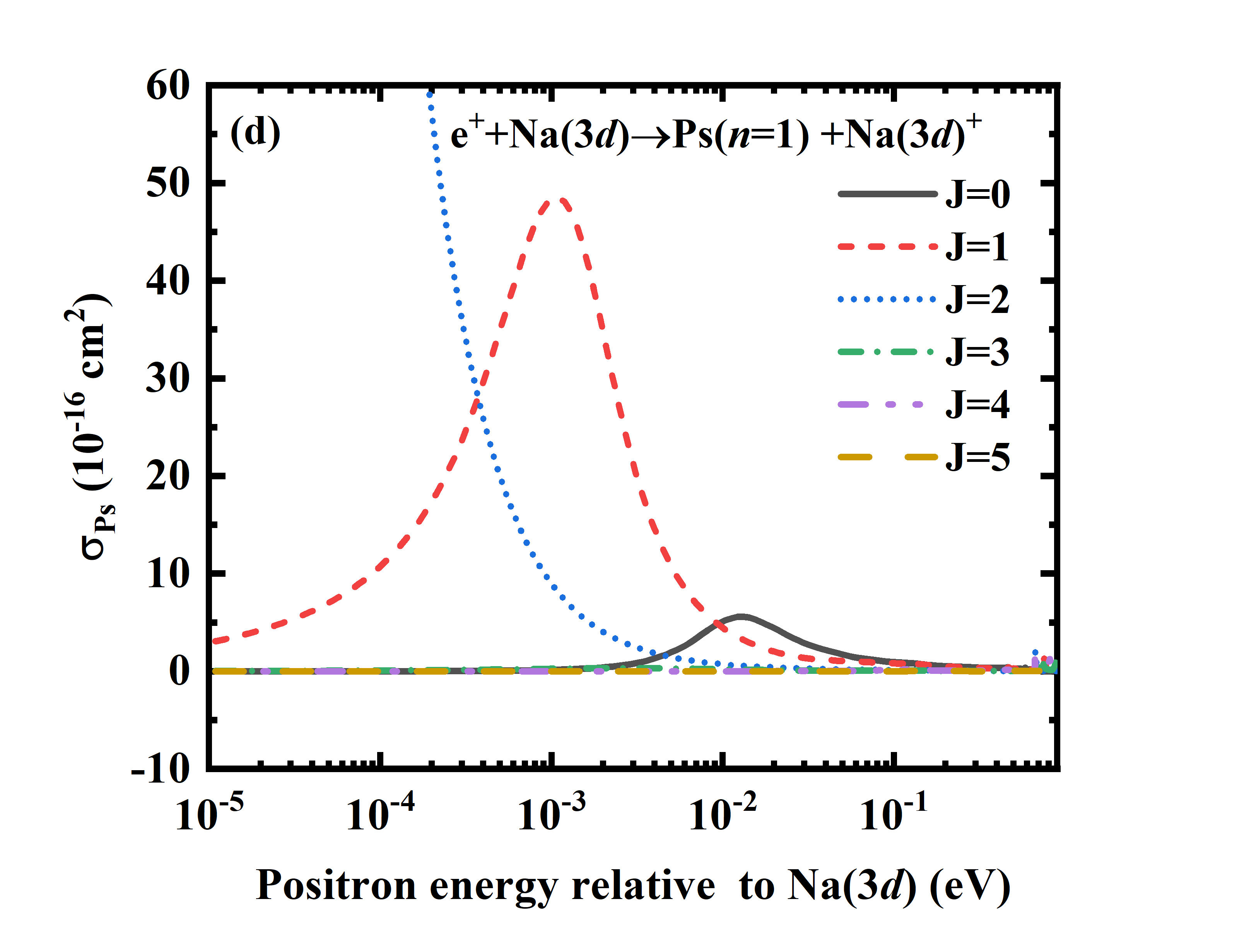}
		\label{fig3d}
	}
	\caption{(Color online) Partial wave cross sections for Ps($n=1$) formation in positron scattering from (a) Na($3s$), (b) Na($3p$), (c) Na($4s$), and (d) Na($3d$) states.}
	\label{fig3}
\end{figure}

Figure~\ref{fig3} presents the partial-wave cross sections for Ps($n=1$) formation in positron scattering from Na($3s$), Na($3p$), Na($4s$), and Na($3d$). It is evident that the primary contributions originate from the partial wave \( J=0 \) for positron scattering from the Na($3s$) and Na($4s$) states, particularly in the very low impact energy region. Notably, the \( J=1 \) and \( J=2 \) partial wave cross sections provide the dominant contributions at low energies for Ps($n=1$) formation in the processes $e^{\scriptscriptstyle+}$ + Na($3p$) $\to$ Ps($n=1$) + Na($3p$)$^{\scriptscriptstyle+}$ and $e^{\scriptscriptstyle+}$ + Na($3d$) $\to$ Ps($n=1$) + Na($3d$)$^{\scriptscriptstyle+}$, respectively. The contributions from higher partial waves are negligible at very low impact energies.

\subsection{Threshold behavior}

We now examine the threshold behavior of cross sections in $e^{\scriptscriptstyle+}$-Na collisions at very low incident positron momenta. Specifically, we investigate the threshold behavior for two processes: (1) elastic scattering, $e^{\scriptscriptstyle+}$ + Na($3s, 3p$) $\to$ $e^{\scriptscriptstyle+}$ + Na($3s, 3p$), and (2) Ps formation, $e^{\scriptscriptstyle+}$ + Na($3s, 3p, 4s, 3d$) $\to$ Ps($n$) + Na($3s, 3p, 4s, 3d$)$^{\scriptscriptstyle+}$, where $n$ is the principal quantum number.

\begin{figure}[htbp]
	\centering
	\subfigure{
		\includegraphics[width=7.8cm]{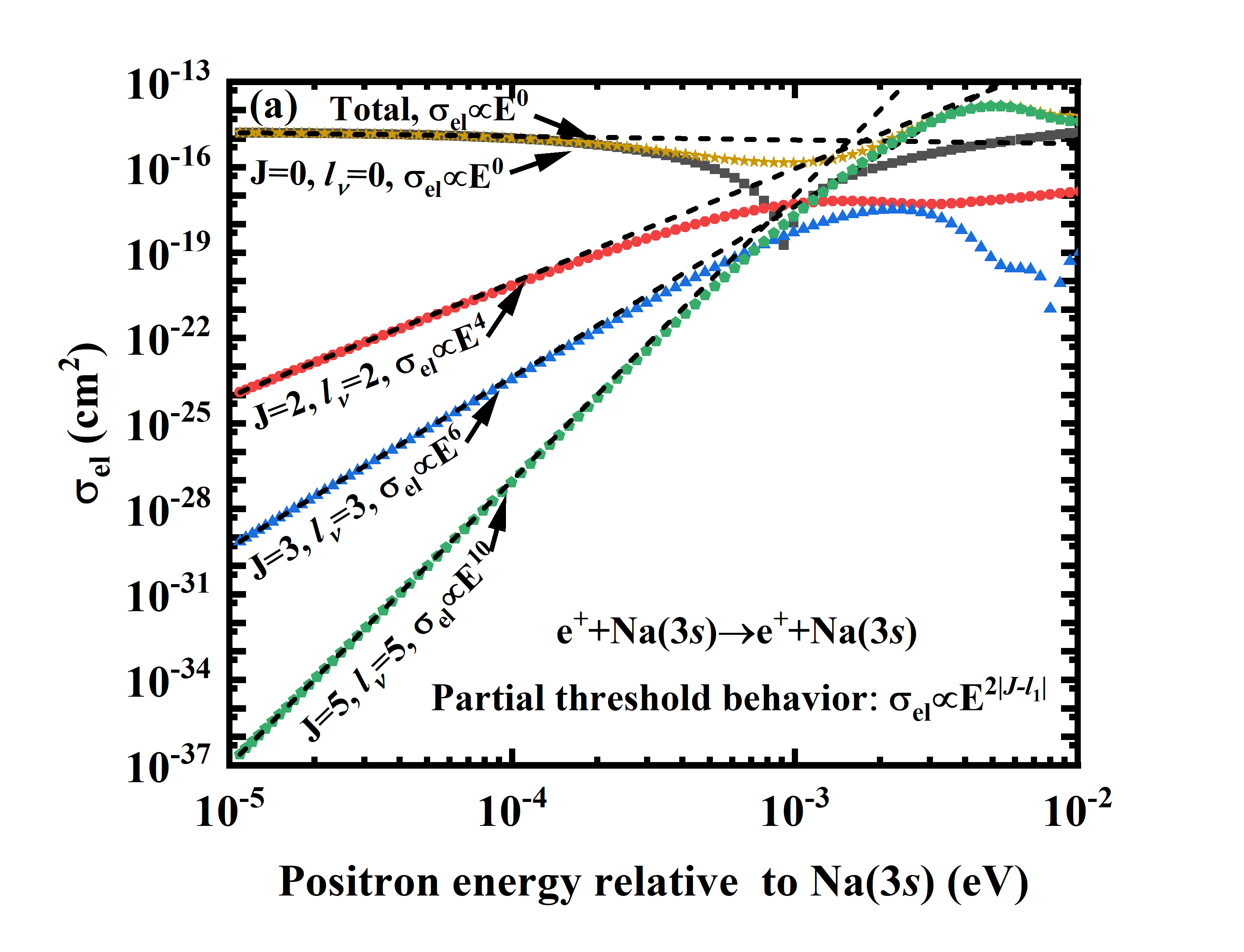}
		\label{fig4a}
	}
	\subfigure{
		\includegraphics[width=7.8cm]{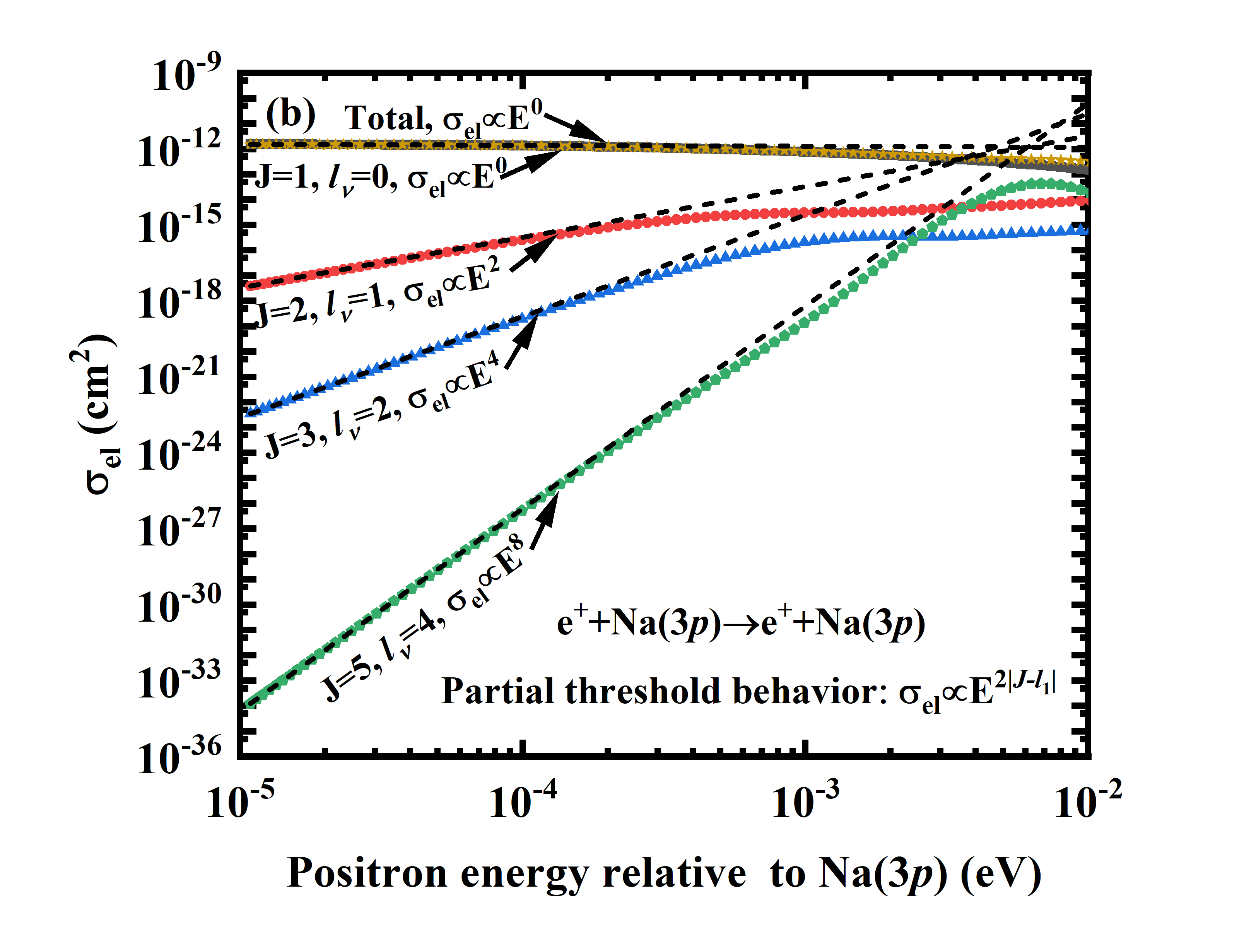}
		\label{fig4b}
	}
	\caption{(Color online) The partial-wave and total elastic cross sections $\sigma_{\text{el}}$ for the collision processes (a) $e^{\scriptscriptstyle+}$ + Na($3s$) $\to$ $e^{\scriptscriptstyle+}$ + Na($3s$) and (b) $e^{\scriptscriptstyle+}$ + Na($3p$) $\to$ $e^{\scriptscriptstyle+}$ + Na($3p$) as a function of the incident positron energy. The threshold behavior for partial waves is indicated by dashed lines.}
	\label{fig4}
\end{figure}

Figure~\ref{fig4} presents the low-energy behavior of the partial-wave and total elastic cross sections $\sigma_{ \text{el}}$ for the processes $e^{\scriptscriptstyle+}$ + Na($3s$) $\to$ $e^{\scriptscriptstyle+}$ + Na($3s$) and $e^{\scriptscriptstyle+}$ + Na($3p$) $\to$ $e^{\scriptscriptstyle+}$ + Na($3p$) as a function of positron energy. As expected, the total $\sigma_{\text{el}}$ approaches a constant value as the positron energy tends to zero, with the primary contribution arising from the zero partial wave at very low energies.

For individual partial waves, our calculations demonstrate that the cross section scales as $\sigma_{\text{el}} \propto E^{2|J-l_{\scriptscriptstyle 1}|}$. This threshold scaling behavior for partial waves is depicted as dashed lines in Fig.~\ref{fig4}. For the process $e^{\scriptscriptstyle+}$ + Na($3p$) $\to$ $e^{\scriptscriptstyle+}$ + Na($3p$), the cross section exhibits an energy dependence similar to that for $e^{\scriptscriptstyle+}$ + Na($3s$) $\to$ $e^{\scriptscriptstyle+}$ + Na($3s$). Notably, the dominant contribution in this case arises from the $J=1$ partial wave.

\begin{figure}[htbp]
	\centering
	\subfigure{
		\includegraphics[width=7.9cm]{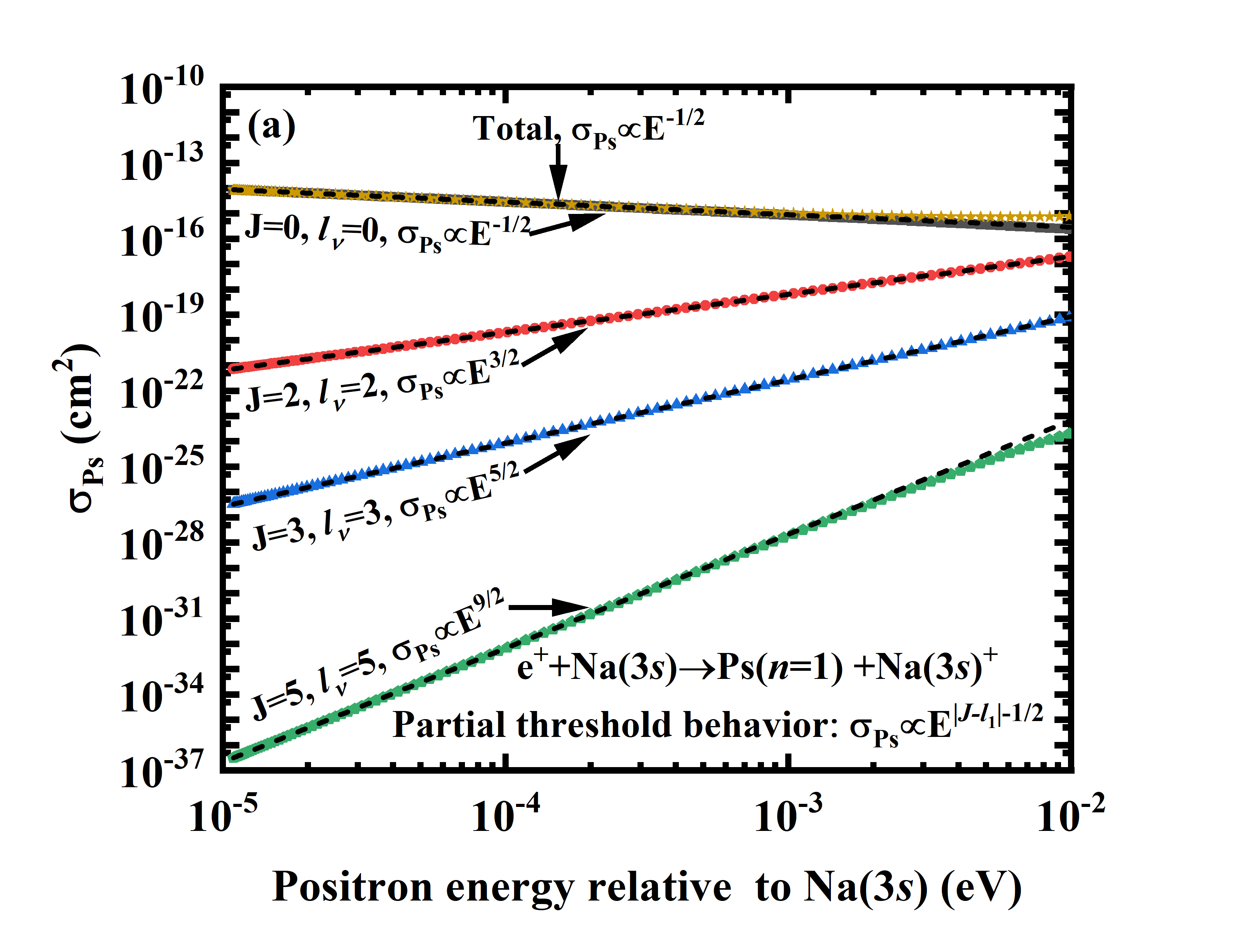}
		\label{fig5a}
	}
	\subfigure{
		\includegraphics[width=7.9cm]{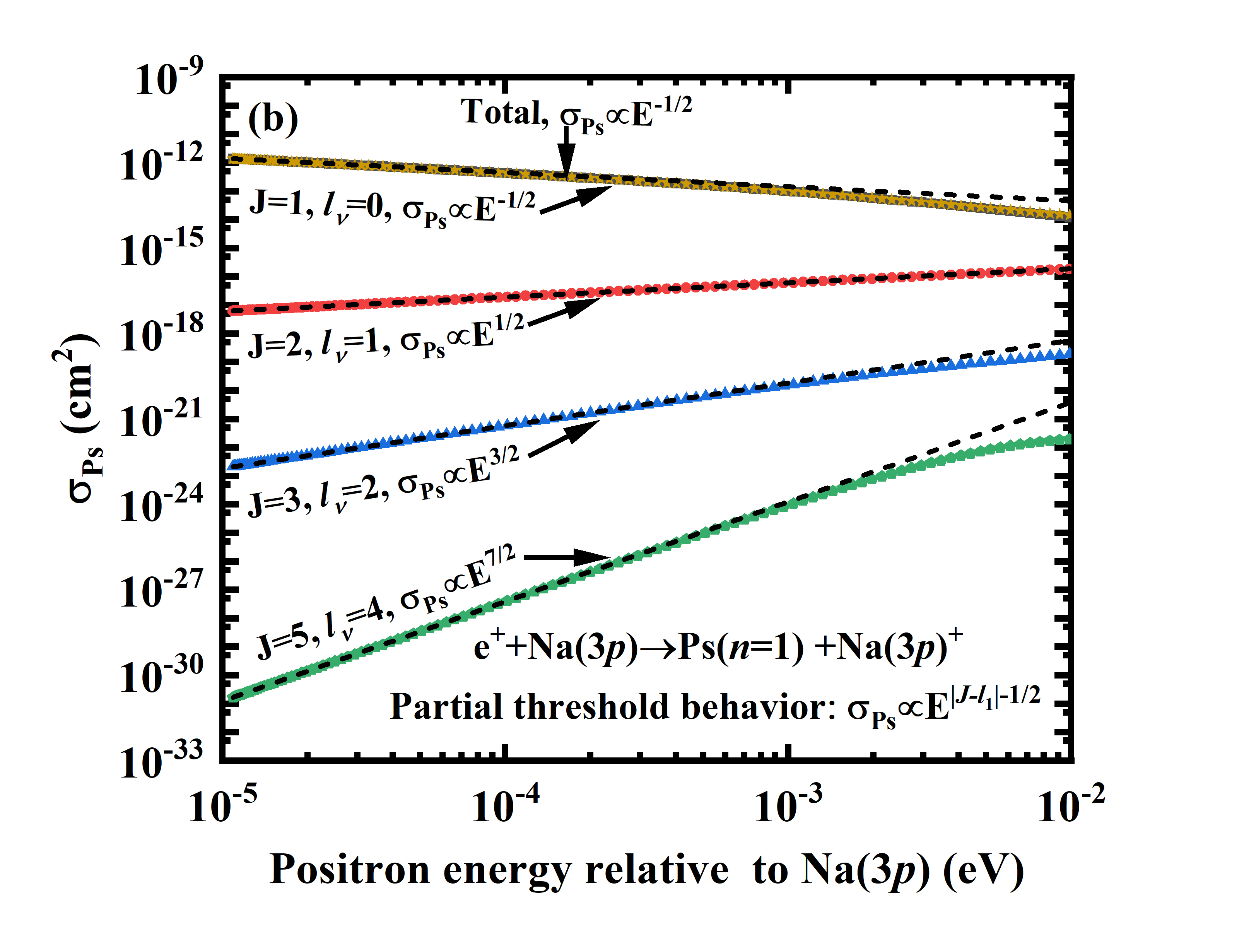}
		\label{fig5b}
	}
\subfigure{
		\includegraphics[width=7.9cm]{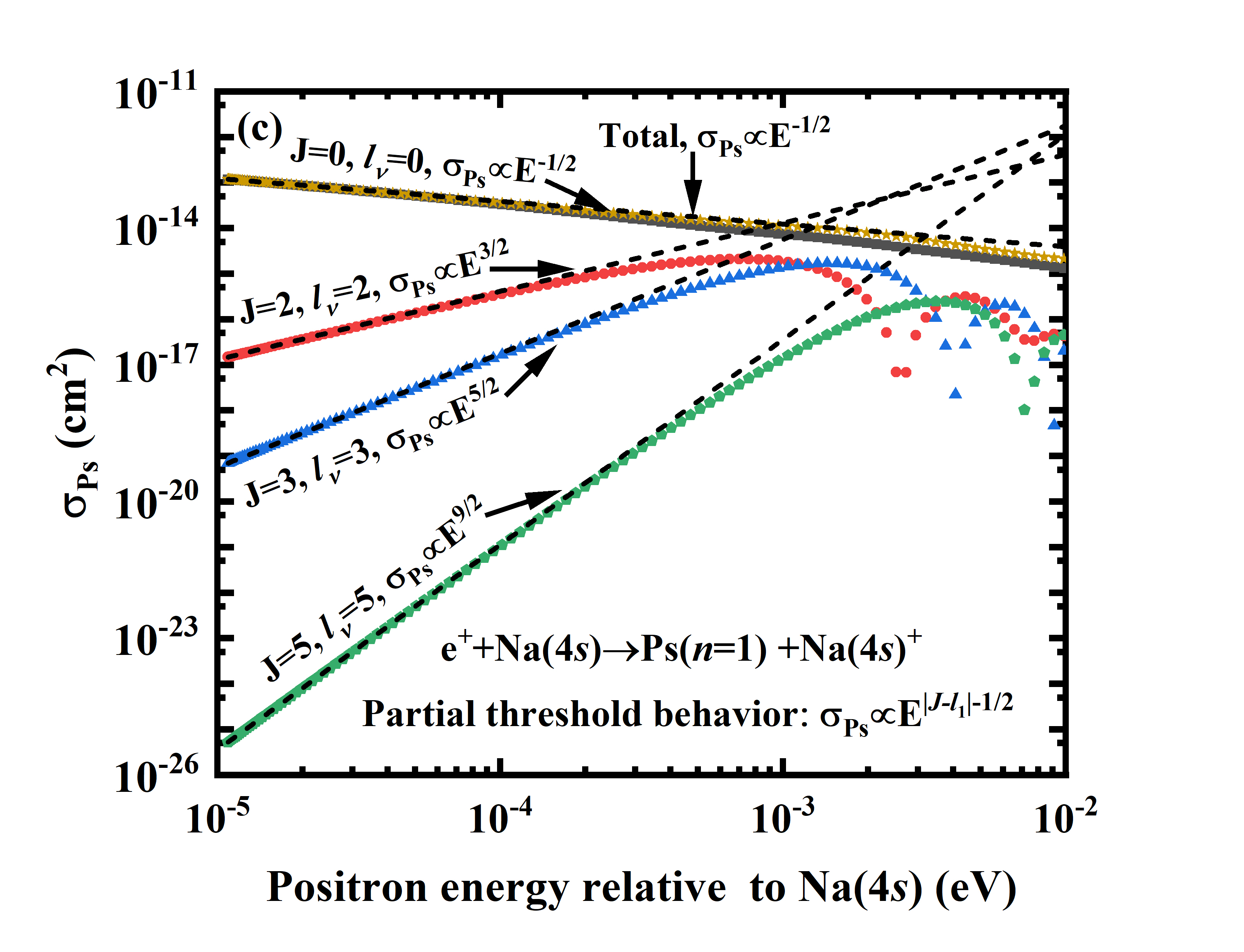}
		\label{fig5c}
	}
\subfigure{
		\includegraphics[width=7.9cm]{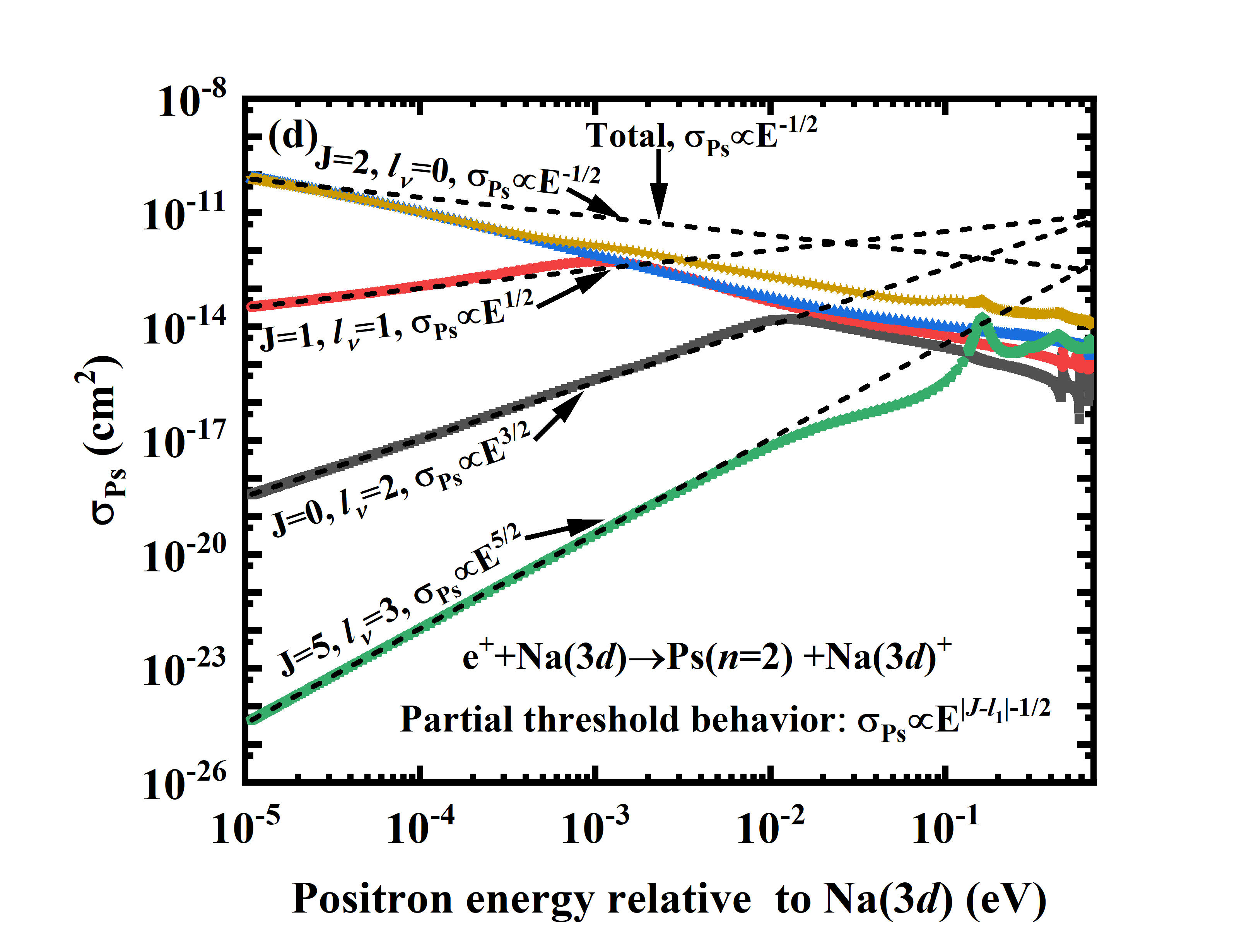}
		\label{fig5d}
	}
	\caption{(Color online) The partial-wave and total threshold behaviors of Ps-formation cross sections are analyzed for several exothermic reaction processes:
 $e^{\scriptscriptstyle+}$ + Na($3s, 3p, 4s$) $\to$ Ps($n=1$) + Na($3s, 3p, 4s$)$^{\scriptscriptstyle+}$ and
$e^{\scriptscriptstyle+}$ + Na($3d$) $\to$ Ps($n=2$) + Na($3d$)$^{\scriptscriptstyle+}$.
The threshold behavior for partial waves is indicated by dashed lines.}
\label{fig5}
\end{figure}

\begin{figure}[htbp]
	\centering
	\subfigure{
		\includegraphics[width=7.8cm]{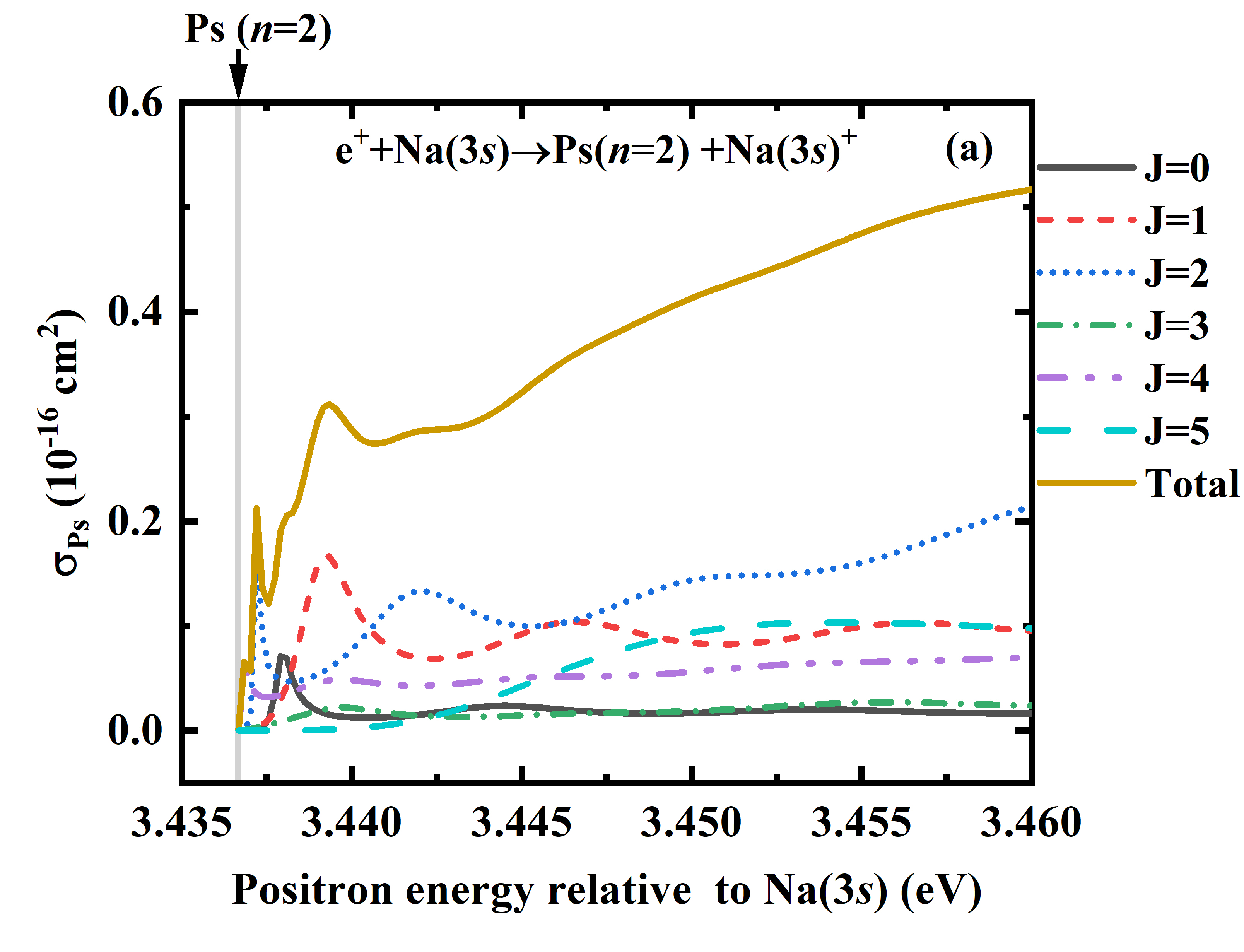}
		\label{fig6a}
	}
	\subfigure{
		\includegraphics[width=7.8cm]{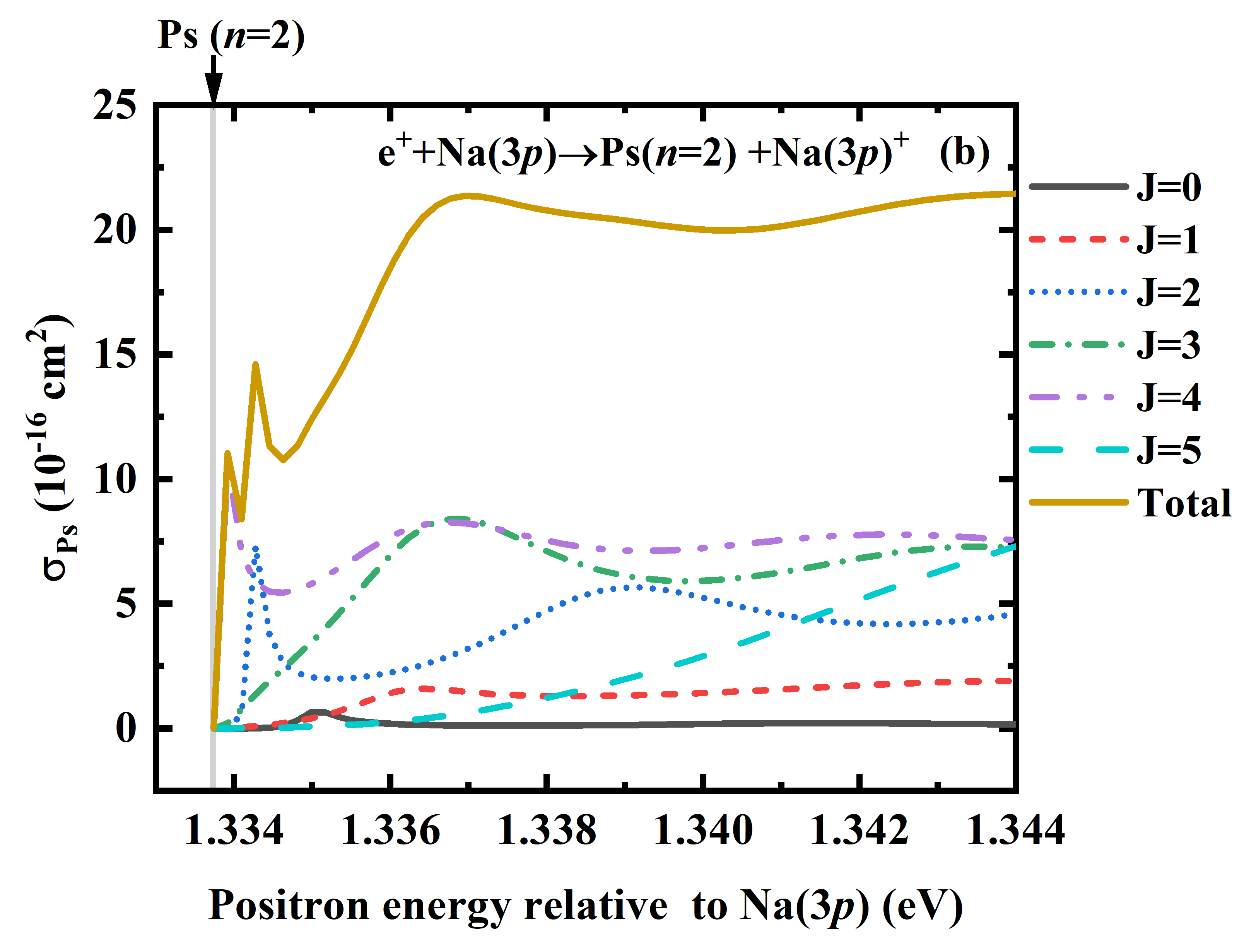}
		\label{fig6b}
	}
\subfigure{
		\includegraphics[width=7.8cm]{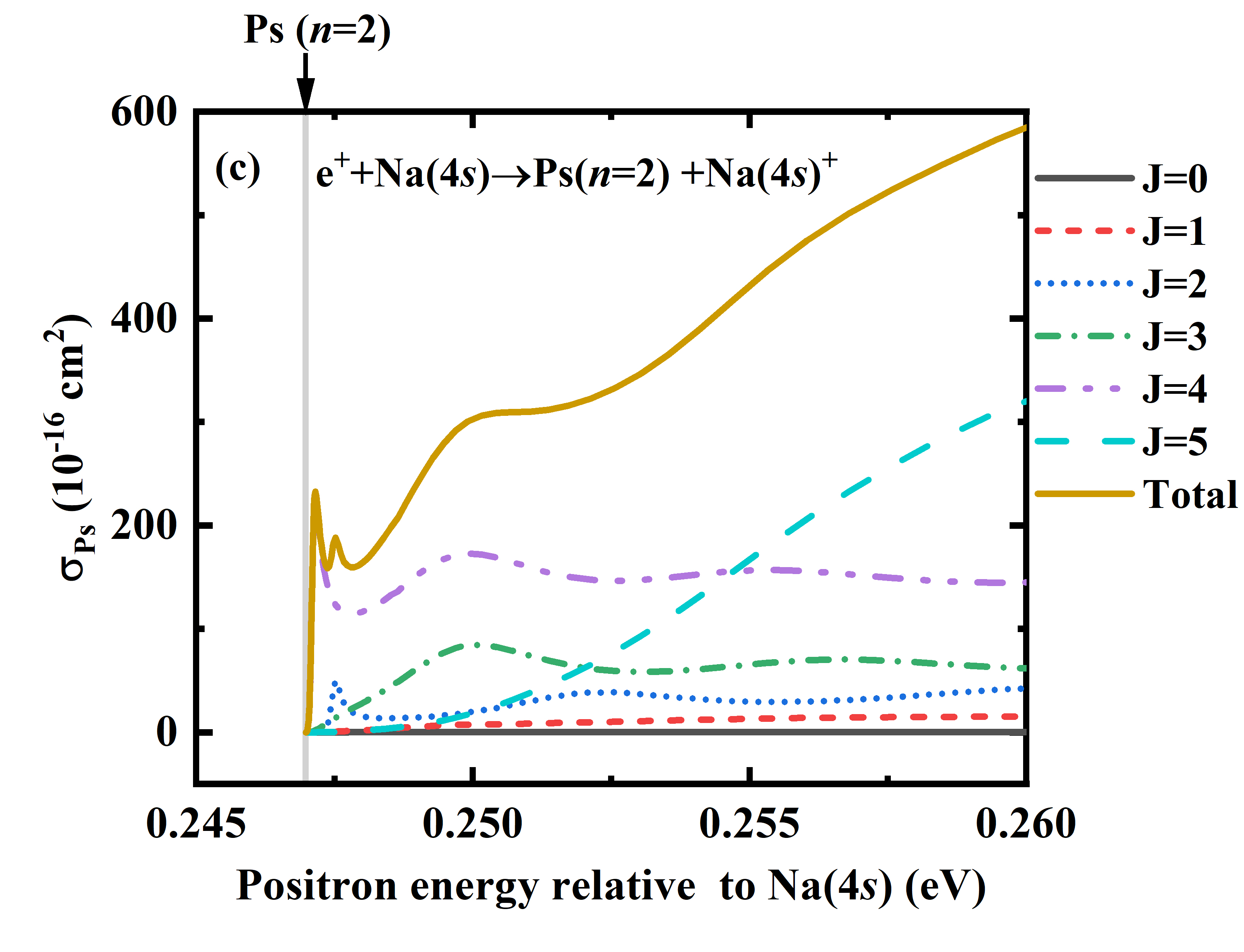}
		\label{fig6c}
	}
	\caption{(Color online) The partial-wave and total threshold behaviors of Ps-formation cross sections are analyzed for several endothermic reaction processes:
$e^{\scriptscriptstyle+}$ + Na($3s, 3p, 4s$) $\to$ Ps($n=2$) + Na($3s, 3p, 4s$)$^{\scriptscriptstyle+}$.
The arrow denotes the position of the Ps($n=2$) threshold.}
\label{fig6}
\end{figure}

Figure~\ref{fig5} presents the partial-wave and total threshold behaviors of Ps-formation cross sections for several exothermic reaction processes:
\begin{align}
e^{\scriptscriptstyle+} + \text{Na}(3s, 3p, 4s) \to \text{Ps}(n=1) + \text{Na}(3s, 3p, 4s)^{\scriptscriptstyle+}\,.
\end{align}
\begin{align}
e^{\scriptscriptstyle+} + \text{Na}(3d) \to \text{Ps}(n=2) + \text{Na}(3d)^{\scriptscriptstyle+}\,.
\end{align}

Our results show that the threshold behavior of Ps-formation cross sections follows the threshold law:
\begin{align}
\sigma_{\text{Ps}} \propto E^{|J-l_{\scriptscriptstyle 1}| - 1/2}\,.
\end{align}
For cross sections summed over all partial waves \( J \), the dominant contribution comes from the initial channel where \( J = l_{1} \). Specifically, for $e^{\scriptscriptstyle+}$ + Na($3p$), the dominant partial wave is \( J = 1 \) , and for $e^{\scriptscriptstyle+}$ + Na($3d$), it is \( J = 2 \). Consequently, the total threshold behavior of Ps-formation cross sections for exothermic reactions follows a
\begin{align}
	\sigma_{\text{Ps}} \propto E^{-1/2}\,.
\end{align}
dependence, leading to a divergence of the cross section at low energies. This trend is illustrated in Fig.~\ref{fig3}. Moreover, our findings align with the threshold behavior of antihydrogen-formation cross sections in exothermic \( \text{Ps}(n_{i},l_{i}) + \bar{p} \) processes, as reported in Ref.\;\cite{hyperfine2018} for the \( J=0 \) partial wave.

For the endothermic process $e^{\scriptscriptstyle+}$ + Na($3s, 3p, 4s$) $\to$ Ps($n=2$) + Na($3s, 3p, 4s$)$^{\scriptscriptstyle+}$, the Ps-formation cross section near the $n=2$ threshold exhibits distinct behavior. While the cross section remains relatively small for $e^{\scriptscriptstyle+}$ + Na($3s$), it increases significantly for $e^{\scriptscriptstyle+}$ + Na($4s$). Just above the Ps($n=2$) threshold, the cross section follows the energy dependence
\begin{align}
	\sigma_{\text{Ps}(n=2)} \propto E^{a}\,,
\end{align}
where \( a \) is positive and depends on the partial wave, as shown in Fig.~\ref{fig6}. As a result, the Ps($n=2$) formation cross section remains finite near the threshold.

\subsection{Comparison with experimental data}
\begin{figure}[htbp]
	\centering
	\subfigure{
		\includegraphics[width=6.7cm]{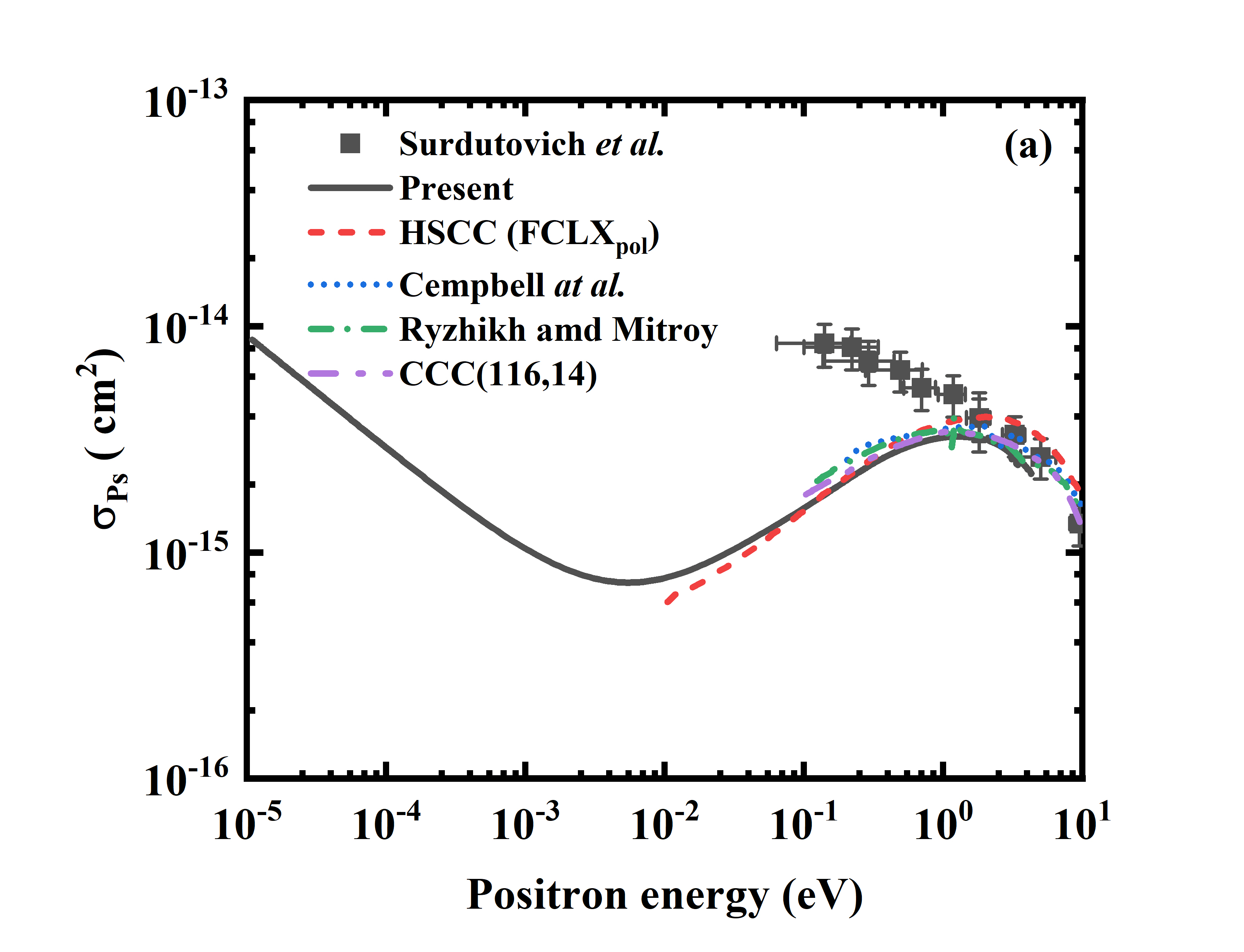}
		\label{fig7a}
	}
	\subfigure{
		\includegraphics[width=9.0cm]{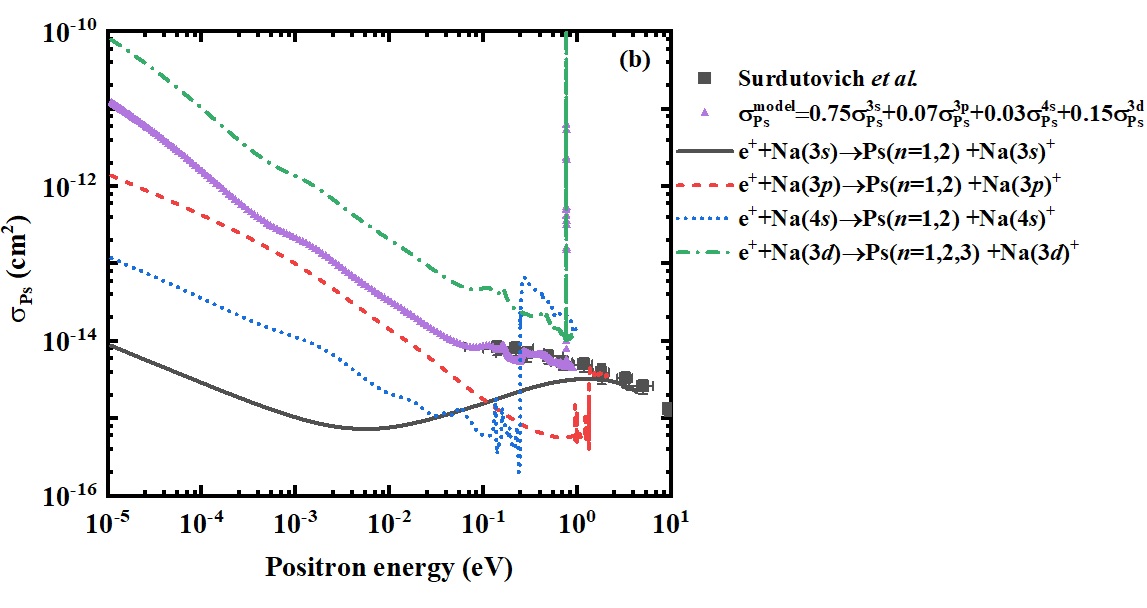}
		\label{fig7b}
	}
	\caption{(Color online) (a) Comparison of Ps-formation cross sections obtained from various theoretical calculations with experimental data from Surdutovich \textit{et al.}\;\cite{surdutovich2002} for the $e^{\scriptscriptstyle+}$ + Na($3s$) $\to$ Ps($n=1,2$) + Na($3s$)$^{\scriptscriptstyle+}$ process. The present results are represented by the solid line. Theoretical calculations include HSCC (FCLX$_{\text{pol}}$) results from \cite{anh2005p}, two CC calculations by Campbell \textit{et al.}\;\cite{campbell1998p} and Ryzhikh and Mitroy\;\cite{ryzhikh1997p}, and the two-center CCC results\;\cite{lugovskoy2012tc}.
(b) Present results of total Ps-formation cross sections were analyzed for several reaction processes:
$e^{\scriptscriptstyle+}$ + Na($3s, 3p, 4s, 3d$) $\to$ Ps($n$) + Na($3s, 3p, 4s, 3d$)$^{\scriptscriptstyle+}$, along with experimental data points from Surdutovich \textit{et al.}\;\cite{surdutovich2002}. The magenta triangles represent the modeled total Ps-formation cross section $\sigma_{\mathrm{Ps}}^{\mathrm{model}}(E)$ obtained by summing contributions from both ground and excited sodium states. 
 }
	\label{fig7}
\end{figure}

Figure~\ref{fig7a} compares the Ps-formation cross sections obtained from various theoretical calculations with the experimental data of Surdutovich \textit{et al.}\;\cite{surdutovich2002}. A key observation is that while theoretical predictions show a decreasing Ps-formation cross section as the positron energy drops below 1 eV, experimental measurements suggest a different trend. Our results for the process $e^{\scriptscriptstyle+}$ + Na($3s$) $\to$ Ps($n=1, 2$) + Na($3s$)$^{\scriptscriptstyle+}$ agree well with other theoretical results in the energy region [0.01, 4.28] eV.

To address the discrepancy between theoretical and experimental Ps-formation cross sections for positron-sodium scattering at collision energies below 1 eV, we analyze Ps formation in positron scattering from the ground-state Na($3s$) atom, followed by scattering from the excited Na($3p$), Na($4s$), and Na($3d$) states, alongside experimental data from Surdutovich \textit{et al.}\;\cite{surdutovich2002}, as shown in Fig.~\ref{fig7b}. We observe that the Ps-formation cross sections generally decrease as the positron energy decreases from 1 eV, then increase and diverge near the threshold. Furthermore, we find that Ps-formation cross sections for positron scattering from excited Na states are significantly larger than those from the ground state in the low-energy region.

Notably, an enhancement in the Ps-formation cross section is observed above the Ps($n=2$) threshold for all initial scattering channels, around \( E = 1 \) eV. The most pronounced increase occurs in the $e^{\scriptscriptstyle+}$ + Na($4s$) and $e^{\scriptscriptstyle+}$ + Na($3d$) processes, beginning at approximately 0.2 eV for $e^{\scriptscriptstyle+}$ + Na($4s$) and above 0.1 eV for $e^{\scriptscriptstyle+}$ + Na($3d$), consistent with the first experimental data point. This enhancement is attributed to Gailitis-Damburg oscillations\;\cite{Gailitis1963}, which result from the long-range dipole interaction between the excited sodium atom and the positron.

Additionally, all theoretical calculations consistently indicate that the Ps-formation cross section does not increase within the energy range of 0.1 to 1\,eV for positron scattering from the ground-state Na($3s$) atom. However, when contributions from excited states are included, the behavior changes significantly. In particular, for excited sodium atoms, the Ps($n=2$) channel opens within this energy range, and Gailitis-Damburg oscillations enhance the Ps-formation cross section, as shown in Figure~\ref{fig7b}. This suggests that the experimentally observed enhancement can be reasonably attributed to the presence of a small fraction of excited sodium atoms in the target.

Although the original experiments did not intentionally excite sodium atoms, it is important to consider possible mechanisms that could lead to their excitation under experimental conditions. Two plausible scenarios are (1) thermal excitation due to the elevated oven temperature and (2) a two-step collision process in which a positron first excites a Na atom (e.g., from $3s$ to $3p$), and a subsequent positron collides with the excited atom to form positronium.

For the thermodynamic case, assuming the Na vapor in the cell follows a Boltzmann distribution, the ratio of excited-state to ground-state population is given by\,\cite{foot2005atomic,bransden2003physics}
\begin{equation}
\frac{n_{\mathrm{exc}}}{n_{\mathrm{ground}}} = \frac{g_{\mathrm{exc}}}{g_{\mathrm{ground}}}\exp\left( -\frac{\Delta E}{k_{B} T} \right),
\end{equation}
Here, $g_{\mathrm{exc}}$ and $g_{\mathrm{ground}}$ are the degeneracies of the corresponding energy levels. $\Delta E = 2.1\,\mathrm{eV}$ is the excitation energy to the $3p$ state, $g_{\mathrm{exc}}/g_{\mathrm{ground}}=3$, and $T = 500\,\mathrm{K}$ (obtained from the saturated vapor pressure of 0.8 mTorr in the Ref\;\cite{surdutovich2002}). This yields $n_{\mathrm{exc}} \approx 2 \times 10^{-21} \, n_{\mathrm{ground}}$, indicating that the thermal population of Na($3p$) is negligible.

To estimate the excited-state number density $P$ resulting from positron-induced collisional excitation, we use the relation\,\cite{bransden2003physics,chenintroduction2016}
\begin{equation}
P = K_{\mathrm{ex}} n_{e^{\scriptscriptstyle+}} n_{\scriptscriptstyle\text{Na}} \tau,
\end{equation}

 Here, $K_{\mathrm{ex}} = \sigma_{\mathrm{exc}} v$ is the excitation rate, where $\sigma_{\mathrm{exc}}$ is the excitation cross section, and $v=k_{ad}/\mu_{ad}$ is the relative velocity between the positron and the sodium atom. $n_{\scriptscriptstyle\text{Na}}$ is the number density of sodium atoms, and $\tau$ is the lifetime of the excited state. The positron number density $n_{e^{\scriptscriptstyle+}}$ is estimated based on a typical slow-positron beam flux of $10^{7}\,\mathrm{cm}^{-2} \mathrm{s}^{-1}$ and a representative beam energy of 2\,eV, corresponding to a positron velocity of $v \approx 8.4 \times 10^{7}\,\mathrm{cm/s}$. This yields a positron number density of approximately $0.12\,\mathrm{cm}^{-3}$. It should be noted that a 90-G axial magnetic field was applied in the experiment, which is expected to increase the effective positron residence time in the interaction region, thereby potentially enhancing the local positron density beyond this conservative estimate.

\begin{figure}[htbp]
	\centering
	\subfigure{
		\includegraphics[width=7.8cm]{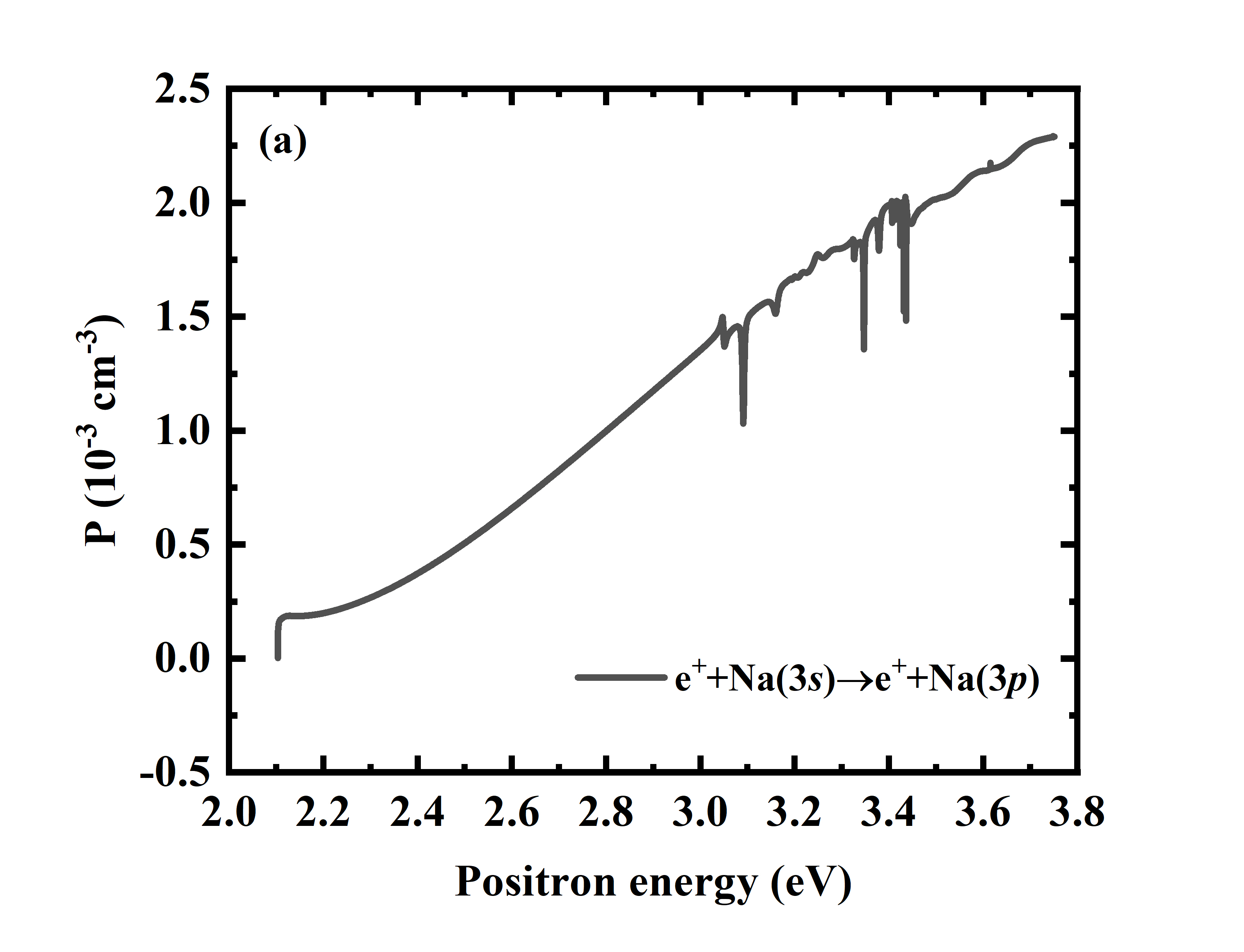}
		\label{fig9a}
	}
	\subfigure{
		\includegraphics[width=7.8cm]{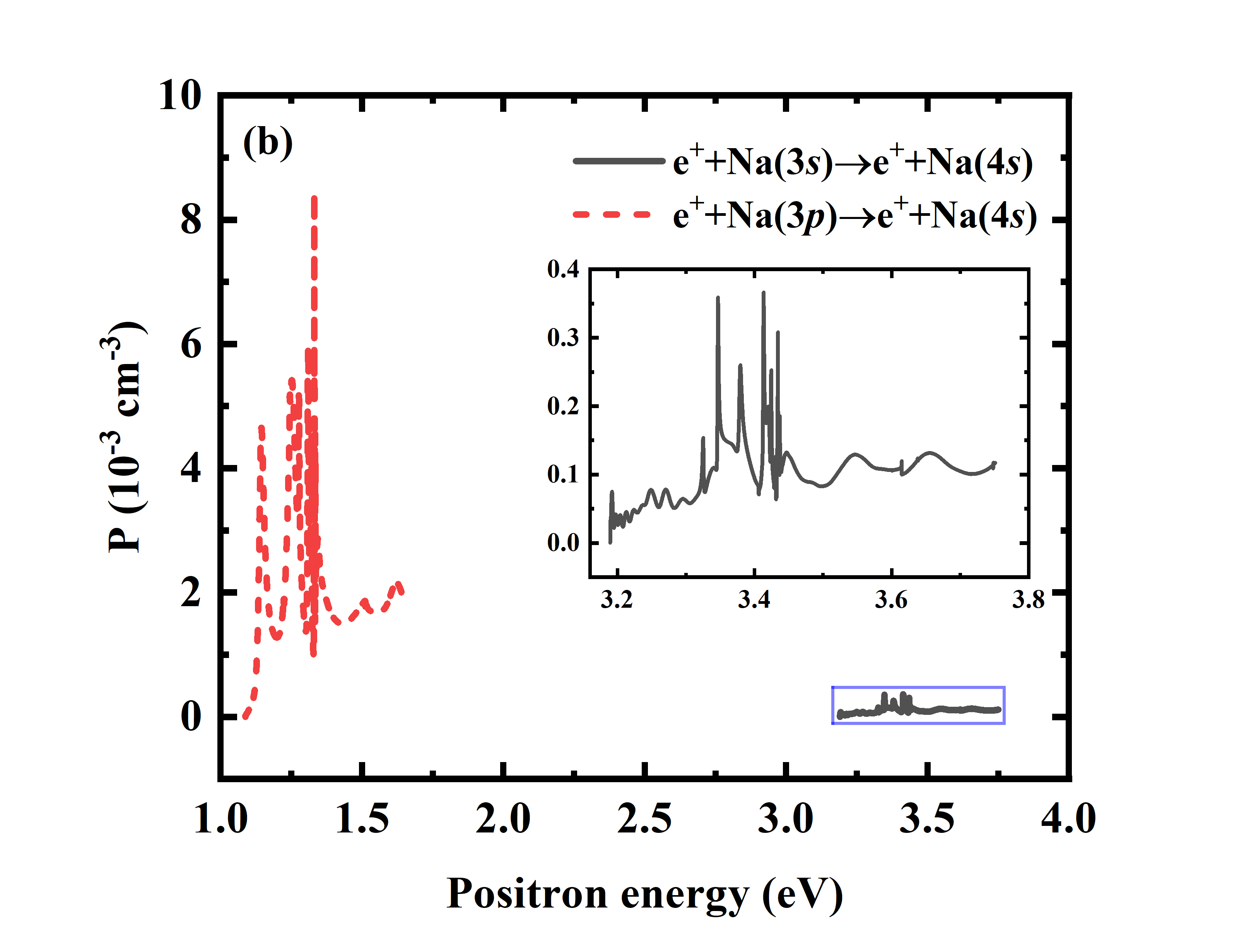}
		\label{fig9b}
	}
	\subfigure{
	\includegraphics[width=7.8cm]{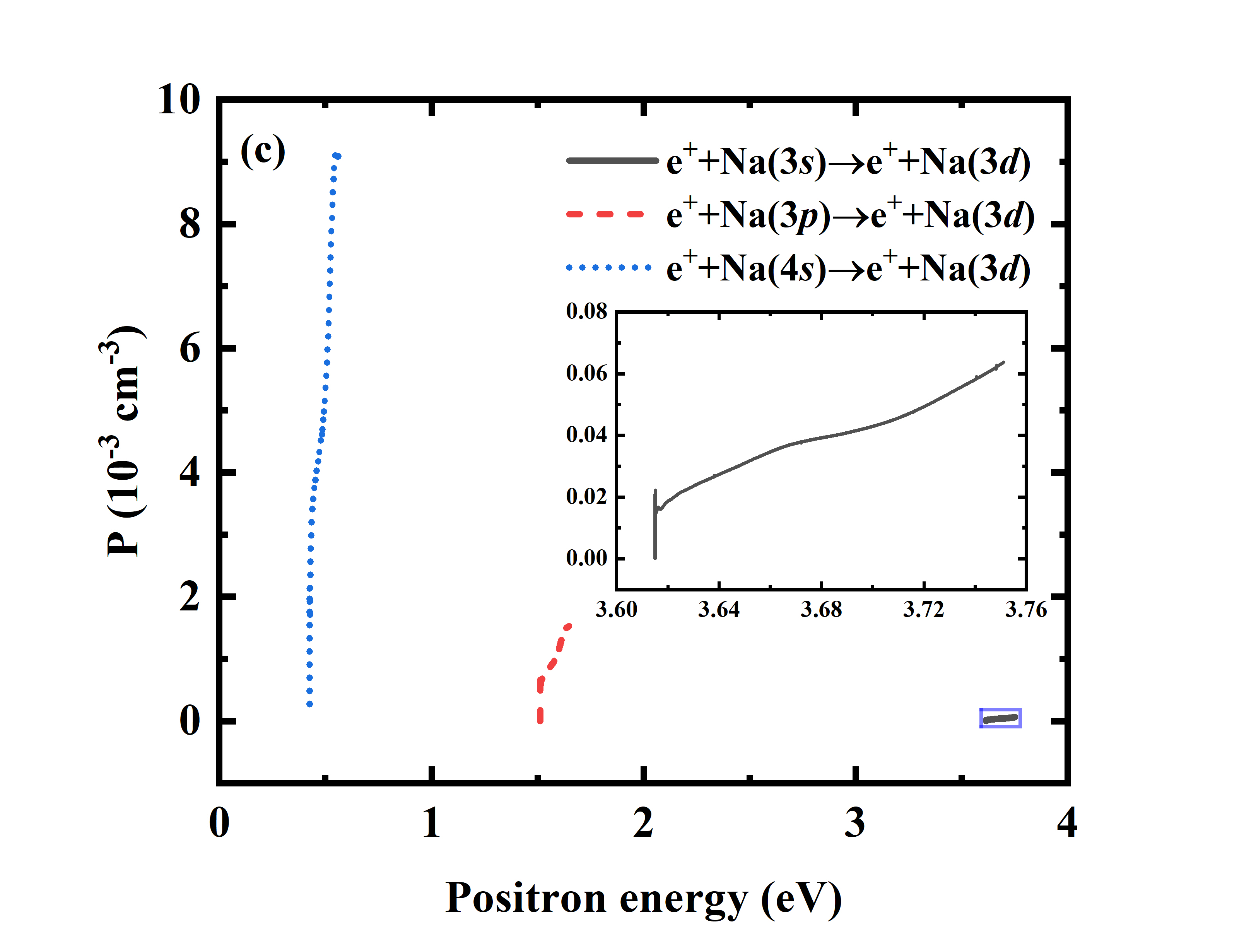}
	\label{fig9c}
}
	\caption{(Color online) The excited-state number density $P$ in the $e^{\scriptscriptstyle+}$-Na system. (a) The excitation process: $e^{\scriptscriptstyle+}$ + Na($3s$) $\to$ $e^{\scriptscriptstyle+}$ + Na($3p$). (b) The excitation processes: $e^{\scriptscriptstyle+}$ + Na($3s$) $\to$ $e^{\scriptscriptstyle+}$ + Na($4s$) (black solid line) and $e^{\scriptscriptstyle+}$ + Na($3p$) $\to$ $e^{\scriptscriptstyle+}$ + Na($4s$) (red dashed line). (c) The excitation processes $e^{\scriptscriptstyle+}$ + Na($3s$) $\to$ $e^{\scriptscriptstyle+}$ + Na($3d$) (black solid line), $e^{\scriptscriptstyle+}$ + Na($3p$) $\to$ $e^{\scriptscriptstyle+}$ + Na($3d$) (red dashed line), and $e^{\scriptscriptstyle+}$ + Na($4s$) $\to$ $e^{\scriptscriptstyle+}$ + Na($3d$) (blue dotted line).}
	\label{fig_exc_prob}
\end{figure}

Using $n_{e^{\scriptscriptstyle+}}\sim 0.12\,\mathrm{cm}^{-3}$, $n_{\scriptscriptstyle\text{Na}} \sim 10^{13}\,\mathrm{cm}^{-3}$, and the calculated excitation rate $K_{\mathrm{ex}}$, we determine the excited-state number density $P$, as shown in Fig.~\ref{fig_exc_prob}. Although the resulting value of $P$ is relatively small (on the order of $10^{-3}\,\mathrm{cm}^{-3}$), it is not negligible, particularly in light of the significantly enhanced positronium formation cross sections associated with excited sodium atoms.

To quantify the impact of excited states on Ps formation, we model the total cross section as a weighted sum over contributions from both ground and excited states:
\begin{equation}
\sigma_{\mathrm{Ps}}^{\mathrm{model}}(E) = f_{3s} \sigma_{\mathrm{Ps}}^{3s}(E) + f_{3p} \sigma_{\mathrm{Ps}}^{3p}(E) + f_{4s} \sigma_{\mathrm{Ps}}^{4s}(E) + f_{3d} \sigma_{\mathrm{Ps}}^{3d}(E),
\end{equation}
where $f_{i}$ are the fractional populations (satisfying $\sum f_{\scriptscriptstyle i} = 1$) and $\sigma_{\mathrm{Ps}}^{i}(E)$ are our computed cross sections for each initial state. By tuning the excited-state fractions (e.g., $f_{3s} = 0.75$, $f_{3p} =0.07$, $f_{ 4s} =0.03$, $f_{3d} = 0.15$), we find that the modeled cross section $\sigma_{\mathrm{Ps}}^{\mathrm{model}}(E)$ closely matches the experimental data, as shown in Figure~\ref{fig7b}. We note that in the fit model, the fractional contribution from Na($3d$) is larger than that from Na($3p$) and Na($4s$). This behavior is consistent with our theoretical cross sections, which show that Ps formation from the $3d$ state is significantly more efficient than that from the $3p$, $4s$ state, particularly at low positron energies. This result demonstrates that including contributions from excited Na states can reproduces the shape of the observed enhancement in Ps formation at low energies.

\section{Summary}

In this paper, we investigated Ps formation ($n=1,2$) and elastic scattering cross sections for the five lowest partial waves in $e^{\scriptscriptstyle+}$-Na collisions. Our analysis was conducted using the $R$-matrix propagation method in the hyperspherical coordinate frame, employing a model potential.

To address the discrepancy between theoretical and experimental Ps-formation cross sections for positron-sodium scattering at collision energies below 1 eV, we examined Ps-formation cross sections for positron scattering from both ground-state and excited-state sodium atoms. We derived an expression for the partial-wave Ps-formation cross section at low positron energies:

\[
\sigma_{\text{Ps}} \propto E^{|J-l_{1}|-1/2}\,.
\]

It is well known that the threshold behavior of inelastic cross sections for two-body systems follows Wigner's threshold law, which predicts \( \sigma_{i\rightarrow j} \propto k_{i}^{2l_{\scriptscriptstyle i} -1} \), where \( k_{i} \) is the initial channel's wave number and \( l_{i} \) is its angular momentum. We found that when the scattered atom is in the ground state, our formula aligns with Wigner's prediction. The reason is that the adiabatic hyperspherical representation reduces the multidimensional Schr$\mathrm{\ddot{o}}$dinger equation to a set of coupled radial equations with an effective angular momentum \( l_{\nu} \), allowing a straightforward application of Wigner's threshold analysis for the ground-state case (\( l_{1} = 0 \)). However, the threshold behavior of positron scattering from an excited-state sodium atom deviates from the standard two-body Wigner analysis. Our formula provides a generalized treatment applicable to both ground-state and excited-state sodium targets.

Our results reveal that the dominant contribution to Ps-formation cross sections at low energies depends on the initial state of the sodium atom. Specifically, the most significant contributions arise when the relative angular momentum between the positron and the sodium atom is zero. Consequently, the total Ps-formation cross section follows the expected threshold behaviors: \( \sigma_{\text{Ps}} \propto E^{-1/2} \) for exothermic reactions, leading to a divergence at low energies. \( \sigma_{ \text{Ps}} \propto E^{a} \) for endothermic reactions, where \( a \) is positive and depends on the partial wave.

We compared our Ps-formation cross sections with both hyperspherical close-coupling (HSCC) calculations and experimental measurements. For positron scattering from ground-state sodium atoms, our results show good agreement with HSCC calculations over the energy range of 0.01 to 4.28 eV. Additionally, our findings indicate that Ps-formation cross sections for positron scattering from excited Na states are significantly larger than those from the ground state in the low-energy region. Prominent Gailitis-Damburg oscillations are observed above the Ps($n=2$) threshold, which may contribute to the observed increase in the experimental data. However, discrepancies between theory and experiment persist when only ground-state sodium targets are considered. Notably, incorporating a small fraction of excited sodium atoms significantly improves the agreement between theoretical predictions and experimental results.

\section{Acknowledgments}

We thank Ke-Dong Wang for helpful discussions. H.-L.H. was supported by the National Natural Science Foundation of China under Grant No. 11874391. T.-Y.S. was supported by the National Natural Science Foundation of China under Grant No.12274423. All the calculations are done on the APM-Theoretical Computing Cluster(APM-TCC).

%

\end{document}